# Uniaxial strain tuning of polar lattice vibrations in KTaO$_3$ and SrTiO$_3$

I. Khayr[1*], N. Somun[2*], S. Hameed[3*], Z. Van Fossan[4], X. He[1], R. Spieker[1], S. Chi[5], E. Clements[5], D. M. Pajerowski[5], M. Minola[3], B. Keimer[3], T. Birol[4], D. Pelc[1,2†], and M. Greven[1†]

[1]School of Physics and Astronomy, University of Minnesota, Minneapolis, MN 55455, USA

[2]Department of Physics, Faculty of Science, University of Zagreb, Bijenička 32, HR-10000 Zagreb, Croatia

[3]Max-Planck-Institute for Solid State Research, Heisenbergstrasse 1, 70569 Stuttgart, Germany

[4]Department of Chemical Engineering and Materials Science, University of Minnesota, Minneapolis, MN 55455, USA

[5]Neutron Scattering Division, Oak Ridge National Laboratory, Oak Ridge, TN 37831, USA

*these authors have contributed equally

†correspondence to dpelc@phy.hr, greven@umn.edu

**The interplay of electronic and structural degrees of freedom is a prominent feature of many quantum materials and of particular interest in systems with strong ferroelectric fluctuations, such as SrTiO$_3$ (STO) and KTaO$_3$ (KTO). Both materials are close to a ferroelectric transition, but despite six decades of extensive research, pivotal questions regarding the nature of this transition and of the associated fluctuations remain debated. Here we combine inelastic neutron scattering, Raman spectroscopy, and *ab initio* calculations to study the evolution of soft polar phonons across the strain-induced ferroelectric transition in STO and KTO. We find that the modes remain underdamped and at nonzero energy, consistent with a first-order quantum phase transition. We also reveal a strong violation of the well-known Lyddane-Sachs-Teller relation between the phonon energies and static dielectric permittivities in insulating KTO and STO, which is not captured by *ab initio* calculations and points to the presence of slow mesoscale fluctuations induced by long-range interactions. In metallic STO, we uncover a first-order transition at a remarkably low critical stress, in qualitative agreement with recent theoretical predictions. The present work resolves several long-standing questions pertaining to the model systems STO and KTO and is relevant to numerous other materials with soft polar phonons.**

Incipient ferroelectrics are of considerable interest both scientifically and with regard to their potential in device applications. In particular, there has been a resurgence of efforts to understand the fascinating properties of the classic model systems STO and KTO [1,2]. The presence of soft polar optic phonon modes leads to large dielectric permittivities in undoped STO and KTO [3,4], yet long-range ferroelectric (FE) order does not appear, and the phonons remain at nonzero energies down to the lowest temperatures [5,6]. The nature of this nearly FE state has been the subject of long-running debate, with theoretical proposals including a quantum fluctuation scenario [7,8], order-disorder effects [9], and fluctuations linked to a flexoelectric interaction between acoustic and optic phonons [10]. Upon doping with electrons, STO displays unconventional superconductivity at extremely low charge carrier concentrations [11,12], and a similar superconducting phase was recently discovered on KTO surfaces and in KTO-containing heterostructures [13,14]. While it is likely that soft polar phonons play a key role in the superconducting pairing [1,2,15-19], the microscopic mechanism remains a major open question.

Due to their proximity to a FE phase, the properties of STO and KTO are sensitive to tiny changes of the crystalline lattice and atomic composition. It is known that long-range FE order can be induced via oxygen isotope substitution in STO [20], and through cation substitution [21-23] and strain [24-34] in both materials. Both epitaxial strain in thin films [28-30] and uniaxial strain in bulk single crystals [24-27,31-33] have been employed to study the nature of this transition and its influence on electronic properties in STO. However, crucial issues remain unresolved. Most importantly, it is not clear if the transition is continuous (second-order) or discontinuous (first-order). The most commonly employed picture is that of a second-order displacive transition, where the energy of the polar phonon mode continuously softens to zero as the critical point is approached [15,21,26,27,34]. For metallic STO, such a second-order quantum phase transition and the associated quantum-critical fluctuations have been proposed to boost the superconducting transition temperature [15,21,35,36]. It has also been predicted, however, that the polar transition is first-order due to an unusual Rashba-like electron-phonon coupling mechanism [37]. Furthermore, evidence of polar nanoregions within the paraelectric phase of both insulating and metallic STO [9,38] has been taken to suggest that the transition has an order-disorder component due to anharmonicities in the local interatomic potential. This would lead to the observed nanoscale polar disorder, and there is evidence that the existence of polar regions causes an enhancement of superconductivity in thin films [29,30]. Yet the soft polar phonon is underdamped in the paraelectric phases of both STO and KTO, which indicates that order-disorder effects are likely not the primary driver of the transition. Determining the evolution of soft phonons across the polar transition is therefore essential to understand the basic physics of both insulating and metallic proximate ferroelectrics, as well as emergent properties such as superconductivity in metallic STO and KTO and related materials.

Here we investigate the behavior of polar soft phonon modes in both insulating STO and KTO, as well as in metallic STO, across the strain-induced FE transition, using inelastic neutron scattering and Raman spectroscopy in combination with specially designed uniaxial strain cells. Bulk elastic strain is a continuous and reversible tuning parameter that has been used in studies of diverse classes of quantum materials – ranging from unconventional superconductors [39-41] to magnetic systems [42,43] – and that is uniquely suited to provide insight into the nature of the polar transition in KTO and STO [24-27]. Neutron scattering enables us to observe the evolution of the soft phonon modes across the transition, and we complement these data with Raman scattering, a highly sensitive probe of polar transverse optic (TO) modes when long-range inversion symmetry is broken in the ordered phase. Generically, the polar phonon is expected to soften continuously to zero energy at a second-order displacive FE quantum phase transition, while its energy should remain nonzero across a first-order transition. In addition, if the transition is of the order-disorder type, phonon overdamping is expected near the transition due to large anharmonicity. Our results are consistent with a first-order transition for all studied samples, and we find that the phonons always remain underdamped. For KTO, our *ab initio* calculations accurately predict the behavior of phonon modes that harden with stress, but do not quantitatively capture the physics of the softening phonons, which indicates that long-range lattice interactions and anharmonicity play an important role.

We apply compressive uniaxial stress along either the [001] or the [1$\bar{1}$0] crystallographic directions of the cubic structure of STO and KTO, as shown in Fig. 1a,b. The neutron scattering measurements employ a high-force pneumatic strain cell (Fig. 1c) that is capable of providing sufficiently high stress levels on large single crystals (see Methods for details), while we use a piezoelectric strain device for Raman scattering (Fig. 1d). Importantly, all neutron measurements are performed in a plane that is perpendicular to the stress and under tensile strain, due to the Poisson ratio; the relevant reciprocal lattice positions are sketched in Fig. 1e. While Raman scattering can only access phonons with nearly zero wavenumber, neutron scattering provides information on the full phonon dispersion. Yet the intrinsic phonon response is convolved with the instrumental resolution function (Fig. 1f,g), which requires a careful data analysis (see Methods).

Figures 2 and 3 show neutron scattering and Raman spectroscopy data for KTO and STO in dependence on compressive uniaxial stress. As noted, we study three distinct systems: insulating KTO and STO, as well as metallic, oxygen-vacancy-doped (OVD) STO with a charge carrier concentration of about $10^{19}$ cm$^{-3}$. KTO is the simplest material of the three, since it remains cubic at all temperatures and only shows a single, doubly-degenerate soft TO phonon mode in the paraelectric phase. The neutron scattering data in Fig. 2a-e show that the phonon properties do not change substantially as the polar transition is approached from the paraelectric phase. For KTO the stress is applied along the [1$\bar{1}$0] crystallographic direction, and we measure phonons that are polarized both along the diagonal [110] direction (Fig. 2a-d and Fig. S2) and along the cubic high-symmetry direction [001] (Fig. 2e). In KTO, the critical stress is accurately known to be 600 MPa from previous dielectric measurements [25], and our Raman data are consistent with this result (Fig. 2f and Fig. S8a). Yet we find that the phonon energies remain nonzero across the polar transition: the [110]-polarized mode hardens (from ~3 to 3.6 meV) throughout the studied stress range (Fig. 2a-d and Fig. S1d-f), while the [001]-polarized mode shows some softening (from ~3 to 2.3 meV) and slight broadening (Fig. 2e and Fig. S1a-c), but does not approach zero energy. Both modes are seen to harden further within the ferroelectric phase, as observed via Raman scattering (Fig. 2h).

In contrast to KTO, STO displays a cubic-to-tetragonal transition near 100 K, and the crystal field anisotropy in the tetragonal phase lifts the phonon degeneracy and splits the soft TO phonon into two modes with distinct energies (Fig. 3a-d). In undoped STO, this splitting is substantial: the energy of the softer $E_u$ mode is 1 meV, while the harder $A_{2u}$ mode resides around 2 meV [44]. Due to their low energies and steep dispersions, it is difficult to resolve the two modes using neutron scattering, and high-resolution cold neutron measurements are needed (Fig. 3a-d). The $E_u$-$A_{2u}$ splitting is expected to decrease with charge carrier concentration [45], so the two modes are again indistinguishable in our metallic STO samples (Fig. 3e). In undoped STO, we apply stress along [001], where the critical strain is much lower than along [110] due to the low energy of the $E_u$ mode [26]. Similar to KTO, we do not find evidence of significant phonon softening; the scattering associated with both $E_u$ and $A_{2u}$ modes is essentially independent of stress in the studied range (Fig. 3a-d, see Figs. S4-7 for additional data). Even though the $E_u$ mode is not resolved in the case of OVD STO, its softening should lead to an increase of the integrated intensity [46] below ~1 meV, which is not observed (Fig. 3d). Fits to the phonon dispersion of undoped STO (Fig. 3a,b; see Methods for details) also do not show appreciable changes of the zone-center energy. As discussed below, prior Raman scattering results on undoped STO [24] match well with the neutron data above the critical stress, so it was not necessary to further extend the neutron measurements.

In metallic STO, the polar phonons are well resolved from the "Bragg tail" in the neutron scattering data (Fig. 3e and Fig. S3) due to the fact that the phonon energy is somewhat higher and the dispersion less steep than in undoped STO and KTO (see Methods for a discussion of resolution issues). The high quality of the data enables us to resolve a subtle initial softening with strain, but the trend is reversed well before the mode energy reaches zero. The upturn appears around 200 MPa, roughly where one of the soft polar TO modes becomes Raman-active (Fig. 3f and Fig. S8b), and it indicates that long-range polar order is established. We note that prior Raman scattering results show evidence of a small volume fraction of the inversion-broken phase already at very small stress, indicated by a nonzero intensity of the hard TO mode at ~170 cm$^{-1}$, which could be due to the presence of dislocations [31], for example. Importantly, the width (or, equivalently, the lifetime) of the soft phonon in both STO and KTO does not change appreciably in the entire stress range (see Fig. S1).

In Fig. 4a, we compare the phonon energies for KTO obtained from the neutron and Raman scattering with our *ab initio* calculation. Since the Raman scattering result was obtained at a higher temperature than the neutron scattering data, we correct the former in order to compensate for the temperature difference (see Methods for details). In Fig. 4a,b, we also calculate the expected phonon energies from prior measurements of the static dielectric constant of undoped KTO [25] and STO [24] using the well-known Lyddane-Sachs-Teller (LST) relation

$$\omega_{TO}^2 = \frac{A}{\varepsilon_0},$$

where $\omega_{TO}$ is the soft TO-mode frequency, $\varepsilon_0$ the static dielectric permittivity, and $A$ a proportionality constant that involves the products of the energies of higher transverse and longitudinal optic (LO) phonons [47] (see also Methods). The LST relation is known to hold well for STO without applied stress [5]. We therefore use the zero-stress energies and $\varepsilon_0$ values to determine $A$ for both STO and KTO. Although we do not directly measure the LO phonons, we find that the TO$_2$ and TO$_4$ phonon energies are stress-independent in strained KTO (Fig. S8c,d). Therefore, we can safely assume that this is also the case for the higher-energy LO phonons and that $A$ is stress-independent. Interestingly, both the *ab initio* calculations (for KTO) and the phonon energies predicted from the LST relation indicate that the [001]-polarized phonons should strongly soften upon approaching the critical stress. Yet this is in stark disagreement with our direct measurements of the phonon energies. Conversely, the stress dependence of the hardening [110]- polarized mode in KTO matches well with both the LST values and *ab initio* calculation. Furthermore, the phonon modes remain underdamped across the transition, *i.e.*, their width is always smaller than their energy.

Our results provide crucial information on the controversial nature of the paraelectric state in KTO and STO, with several important implications. First, the strain-induced polar transition in these representative incipient ferroelectrics is clearly not second-order, *i.e.*, the relevant phonon modes do not soften toward zero energy. We note that evidence for a second-order transition in strained STO membranes has been obtained from x-ray absorption spectroscopy [34], but this technique cannot distinguish between a spatially homogeneous increase of the polar order parameter and changes in the ordered volume fraction. Both our *ab initio* calculation and previous mean-field theoretical analysis [26] assume a second-order transition, but these approaches do not adequately consider long-range fluctuations and nonlinear effects. Generically, it has been shown that structural phase transitions that involve the softening of an optic phonon branch become first-order due to the Larkin-Pikin mechanism, *i.e.*, through interactions between optic and long-wavelength acoustic modes [48]. Hybridization between optic and acoustic phonons is well documented in both STO [49] and KTO [50], and the Larkin-Pikin effect could preclude a second-order transition with stress as well. Furthermore, undoped STO and KTO show appreciable softening of the transverse acoustic branch upon cooling and it has been argued that an incommensurate ordered phase might form as a precursor to the homogeneous polar phase [51,50]. Yet this would lead to the appearance of additional Bragg peaks with increasing stress, which we do not observe within our experimental sensitivity. Our findings are therefore in qualitative agreement with a recently proposed scenario whereby fluctuations destroy the long-range incommensurate order [10], and it would be interesting to perform a targeted search for diffuse scattering signatures of short-range correlations that have been predicted to be associated with this effect.

In metallic STO, the evolution of the phonon parameters with stress is particularly clear, and the critical stress at which the polar order sets in is remarkably low. The zero-stress TO phonon energy in our metallic STO sample is nearly the same as for KTO, but the critical stress is about three times smaller (~200 MPa vs 600 MPa, respectively). This is in qualitative agreement with recent theoretical work on metallic systems that are close to an inversion-breaking transition, where the dominant electron-phonon coupling is an effective dynamic Rashba interaction [37]. The nominally second-order polar transition is then predicted to be preceded by a first-order transition at smaller values of the control parameter (strain, in our case). A simple linear extrapolation of the stress dependence of $\omega_{TO}^2$ yields a critical stress of ~600 MPa for the underlying second-order transition, similar to the value for undoped KTO. The increase of superconducting transition temperatures in strained STO [52] is therefore likely unrelated to phonon softening, but rather a consequence of the emergence of polar order, which is known to boost superconductivity [29,30] and has been explained within a recent strong-coupling theoretical analysis of superconducting pairing caused by the dynamic Rashba interaction [18]. The present measurements of the phonon energies across the polar transition provide key information for further detailed tests of this theory.

We conclude with a number of observations. The observation of underdamped phonon modes and the concomitant LST breakdown are unusual: the LST relation relies on homogeneity and basic phonon electrostatics, so its breakdown implies that the system must become inhomogeneous close to the transition, but in such a way that the phonon energies and linewidths are not substantially affected. First-order FE transitions are associated with nucleation and growth of slowly fluctuating polar regions that cause the increase of the low-frequency permittivity. Yet this also leads to nanoscale inhomogeneity that should broaden the phonon linewidths, as observed in other ferroelectrics such as $BaTiO_3$ [53]. The absence of a strong broadening indicates that mesoscale ordered regions appear at sufficiently large length scales such that the TO phonons remain locally well-defined. Mesoscale polar domain structures have been associated with superconductivity enhancement [31,32] and emergent magnetism [54] in plastically deformed STO, and targeted studies of such effects will be highly interesting in STO, KTO, and other materials that show prominent soft polar phonon modes, such as (Pb,Sn)Te [55] and the multiferroic $EuTiO_3$ [56]. Our findings of LST breakdown and of a robust first-order phase transition in both STO and KTO have broader implications for incipient ferroelectrics and quantum criticality – the results for these two model systems indicate that it is difficult to realise a true displacive second-order FE quantum phase transition, since it can be cut off either by long-range elastic interactions or dynamic Rashba coupling of electrons to the soft polar phonon. Finally, we note that the capability of inelastic neutron scattering measurements with high-force strain cells demonstrated in the present work opens possibilities for strain tuning of both lattice and magnetic excitations in a wide range of quantum materials.

**Methods**

*Sample preparation.* The STO and KTO crystals used in this study were obtained commercially (MTI Corporation, Princeton Scientific, BIOTAIN Inc., SurfaceNet) and polished to high precision for the *in situ* strain measurements. The neutron scattering samples have masses of 3.4 g (undoped STO), 0.9 g (KTO), 2 g (OVD STO 1) and 0.8 g (OVD STO 2). The metallic STO samples were electron-doped via the introduction of oxygen vacancies by annealing in high vacuum at 980°C for 24 hours. The carrier concentration was determined from measurements of quantum oscillations using a contactless method, with the sample placed in a coil that is part of a resonant LC circuit in the radio frequency range. Highly sensitive measurements of the resonant frequency using a vector network analyzer were then used to extract the resistivity of the sample in dependence on external magnetic field. Clear quantum oscillations were observed at 1.5 K in fields up to 12 T, and a comparison with prior studies [57] enabled us to estimate the carrier density to be approximately $1.5 \cdot 10^{19}$ cm$^{-3}$, where two electronic bands cross the Fermi level. This value also agrees well with the known shift of the phonon energy with respect to undoped STO [5].

*Neutron scattering.* Inelastic neutron scattering measurements were performed at Oak Ridge National Laboratory, with the HB-3 triple-axis spectrometer at the High Flux Isotope Reactor (KTO and OVD STO) and the cold neutron chopper spectrometer (CNCS) [58] at the Spallation Neutron Source (undoped STO). The triple-axis experiments used collimations 48"-40"-sample-40"-120" and a final energy of 13.5 meV, with a full-with-at-half-maximum energy resolution slightly below 1 meV at the elastic line, as calibrated with a vanadium standard. The CNCS measurements were performed with incident energies of 8.5 meV and 12 meV, and a chopper frequency of 300 Hz (high-flux setting). The scattering associated with soft TO phonon modes was measured around the (200) Bragg position. We used liquid helium evaporation cryostats with a base temperature of 1.5 K at both instruments.

For *in situ* uniaxial stress, we employed a custom-made pneumatic strain cell with a maximum force on the sample of 10 kN perpendicular to the scattering plane, continuously tunable through changes of the helium gas driving pressure. The cell is conceptually similar to devices used previously in elastic strain measurements of magnetic rare-earth titanates [42] and in studies of plastically deformed STO and KTO [31,32], but with a larger piston diameter and maximized free angular range in the horizontal scattering plane. In order to decrease background scattering, the cell components were carefully shielded with cadmium foil (CNCS) or gadolinium oxide sheets (HB-3). The sample deformation was measured independently with an inductive sensor (variable differential transformer). Stress homogeneity was maximized through the use of samples with a high length-to-width ratio and of anvils with surface areas similar to those of the samples. We did not observe significant deviations from homogeneity up to the highest stress values: the Bragg peak width remains unchanged within the instrumental resolution, and the TA modes in KTO remain sharp.

*Neutron scattering data analysis.* In triple-axis spectroscopy, the intrinsic sample response is convolved with the instrumental resolution, which takes the form of an ellipsoid in reciprocal and energy-transfer space. In order to extract the phonon parameters, we corrected the raw data for the bosonic occupation (Bose factor) and performed fits of damped harmonic oscillator (DHO) susceptibilities convolved with the HB-3 resolution using the MATLAB library ResLib [59]. The fact that the ellipsoid integrates over a portion of the phonon dispersion was taken into account by including the known low-energy dispersion of the TO modes in the DHO susceptibilities [48] (see also below).

In the case of CNCS, the reciprocal-space resolution is determined by a combination of the size of the data bins and the intrinsic instrument resolution, while the energy resolution is known for a given instrument configuration [58]. For the zone-center cuts in Fig. 2c,d, acceptable statistics was obtained with a bin size of ±0.02 r.l.u. in each reciprocal space direction, which caused some artificial broadening due to integration over the dispersion. The energies and wavenumbers of the TO $A_{2u}$ and $E_u$ modes in Fig. 3d,e were obtained from Lorentzian fits to constant-energy slices of the data with 8.5 meV incident energy, assuming a stress-independent amplitude ratio for the two modes; fits to the raw data are shown in Fig. S4. Constant-energy slices were used, because in this case the phonons are better resolved than in constant-energy cuts due to the relatively steep dispersion. For fits to the TO dispersions (Fig. 3a,b), we use the simplest quadratic form,

$$\omega_{1,2}^2(q) = \omega_{A2u,Eu}^2 + v_{TO}^2 q^2$$

where $\omega_{A2u,Eu}$ are the zone-center energies of the two modes, and $v_{TO}$ is the velocity parameter. In order to minimize the number of free parameters, we took $v_{TO}$ to be the same for both modes and independent of stress; both assumptions are borne out from the data with incident energy 12 meV, where a wider energy and $q$ range is covered, but the relatively poor energy resolution does not permit a full analysis of the low-energy part of the dispersion (Figs. S4-6). We note that similar results were obtained if $v_{TO}$ is left as a free parameter for each stress.

*Raman scattering.* For the *in situ* stress measurements, single crystals were cut and polished into a needle shape with approximate dimensions of 2.5 mm (length) × 200 μm (width) × 100 μm (thickness). The samples were then clamped onto a sample carrier using Stycast 2850FT epoxy, which was subsequently attached to a piezoelectric strain device. Stress was controlled by applying a voltage to the piezoelectric stacks.

The Raman scattering experiments were performed in backscattering geometry using a Jobin-Yvon LabRAM HR800 spectrometer from HORIBA. A HeNe laser with a wavelength of 632.8 nm was used. The laser beam spot (diameter ~ 2 μm) was significantly smaller than the exposed strained area of the sample (~600 × 200 μm²), which ensured uniform strain in the probed volume. A closed-cycle cryostat was used for cooling, and all measurements were performed at its base temperature of $T$ = 30 K. A diffraction grating with 1800 grooves/mm was employed, yielding a spectral resolution of ~1.5 cm$^{-1}$. In order to minimize variations due to strain inhomogeneity, Raman spectra were collected from the same position on the sample throughout the experiment. For [001] strain on OVD-STO, measurements were performed with the incident photons propagating along [100] and polarization along [010], perpendicular to the strain direction. For [1$\bar{1}$0] and [001] strain in KTO, the incident photons propagated along [001] with along [1$\bar{1}$0] - parallel and at 45° relative to the strain direction, respectively. In all cases, the outgoing polarization was not analyzed.

For details pertaining to stress calibration and additional Raman scattering data, see Figs. S9-12 and the corresponding sections of the Supplemental Material.

*LST relation and ab initio calculation.* The full LST relation for an insulator with multiple phonon branches reads [5]

$$\frac{\varepsilon_0}{\varepsilon_\infty} = \frac{\prod_i \omega_{LO,i}^2}{\prod_i \omega_{TO,i}^2}$$

where $\varepsilon_0$ and $\varepsilon_\infty$ are the static and high-frequency dielectric permittivities, respectively, and $\omega_{TO,i}$ and $\omega_{LO,i}$ are the zone-center TO and LO phonon energies of the *i*-th branch. The relation in the main text combines $\varepsilon_\infty$ and the higher TO and LO mode energies into the constant *A*. We note that a generalized version of the LST relation that includes the imaginary part of the longitudinal frequencies (*i.e.*, damping) was derived long ago [60]. The frequencies $\omega_{TO,i}$ then correspond to the DHO frequencies, and $\omega_{LO,i}$ are replaced by absolute values. This modification, however, does not impact our analysis, especially for KTO: the longitudinal modes are not substantially damped, and we obtain the TO frequencies from DHO fits and not from the positions of the susceptibility peaks.

First-principles density functional theory calculations were performed with the Vienna Ab-initio Simulation Package (VASP) using the PBEsol exchange-correlation functional to calculate the soft polar phonon frequency of uniaxially strained $KTaO_3$ [61-63]. The effects of uniaxial strain were simulated by relaxing lattice vectors and atomic positions along the *x-y* plane while holding their *z*-coordinates fixed. Strain was then imposed along different crystallographic directions by reorienting the unit cell such that the direction of interest was parallel to the *z*-axis. Relaxations and phonon calculations were performed with 8x8x8 and 12x12x12 *k*-point meshes respectively with an energy cutoff of 600 eV and spin-orbit coupling (SOC) accounted for. In order to align with experimentally reported values for the soft phonon frequency in bulk KTO, a hydrostatic pressure of -20 kbar was imposed in all unit-cell relaxations. Further discussion of the imposed hydrostatic pressure and additional phonon frequency calculations are provided in the Supplementary Information (Fig. S13).

**Acknowledgements**

We thank Chris Leighton, Marin Lukas, Maria Navarro Gastiasoro and Avraham Klein for helpful discussions and comments; Arthur von Ungern-Sternberg Schwark, Mark Joachim Graf von Westarp, and Lichen Wang for technical assistance with Raman and XRD measurements performed under strain; and Marin Lukas and Clifford Hicks for the design and development of the pneumatic and piezoelectric uniaxial strain devices, respectively. Work at the University of Minnesota was funded by the U.S. Department of Energy through the University of Minnesota Center for Quantum Materials, under Grant No. DE-SC-0016371. Work at the University of Zagreb was funded by the Croatian Science Foundation under Grant No. UIP-2020-02-9494 and the Croatian Ministry of Science, Education, and Youth. A portion of this research used resources at the Spallation Neutron Source and High Flux Isotope Reactor, a DOE Office of Science User Facility operated by the Oak Ridge National Laboratory. The beam times were allocated to CNCS on proposal IPTS-26911 and HB-3 on proposals IPTS-31592 and IPTS-33290.

**Author contributions**

DP and MG conceived the research. IK, NS, DP, RS, XH, SC, EC and DMP performed neutron scattering experiments and analysed data, guided by MG and DP. SH and MM performed Raman scattering experiments and analysed data, supervised by BK. ZVF and TB performed *ab initio* calculations. DP, SH, IK, NS and MG wrote the paper, with input from all authors.

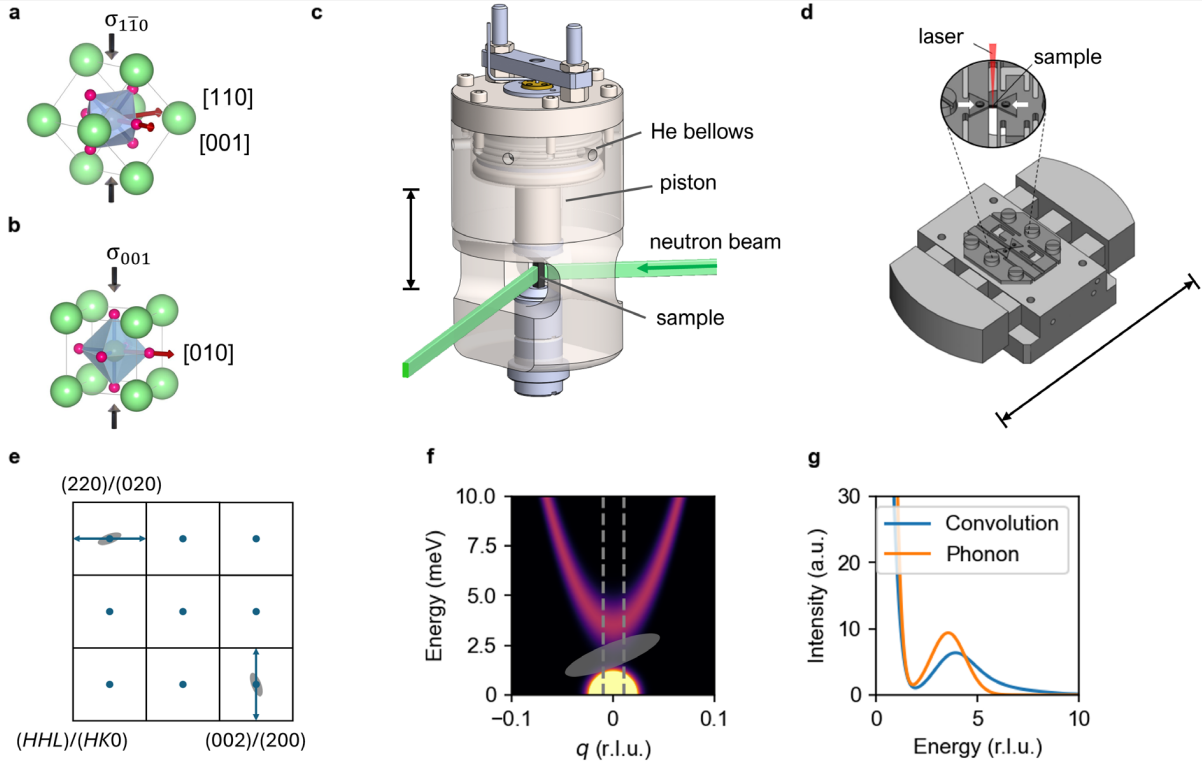

**Figure 1 | Schematics of crystal structures, strain cells, and neutron scattering scans.** Crystal structures of (**a**) KTO and (**b**) STO with stress orientations and phonon polarizations. Schematic of (**c**) the neutron scattering pneumatic strain cell and (**d**) Raman scattering piezoelectric strain device with their respective scales. The scale bars are both 5 cm in length. (**e**) Reciprocal-space map of the Brillouin zones that were investigated in this study. In KTO, stress was applied along [1$\bar{1}$0], restricting reciprocal-space probing to the *HHL* plane. In STO, the *HK*0 plane was probed with stress applied along [001]. A projection of the resolution ellipsoid (grey oval) is indicated in the Brillouin zones of interest. (**f**) Momentum- and energy-resolved intensity map of a simulated phonon at the Brillouin zone center, with the projected resolution ellipsoid comparable to the neutron scattering setup. As shown in (**g**), an energy scan (with a nonzero *q* integration window represented by the two grey dashed lines in (**f**)) through an underlying phonon with energy $\omega_{TO}$ (orange) results in an effective intensity peak (blue) at a higher energy due to the nonzero extent of the instrument resolution.

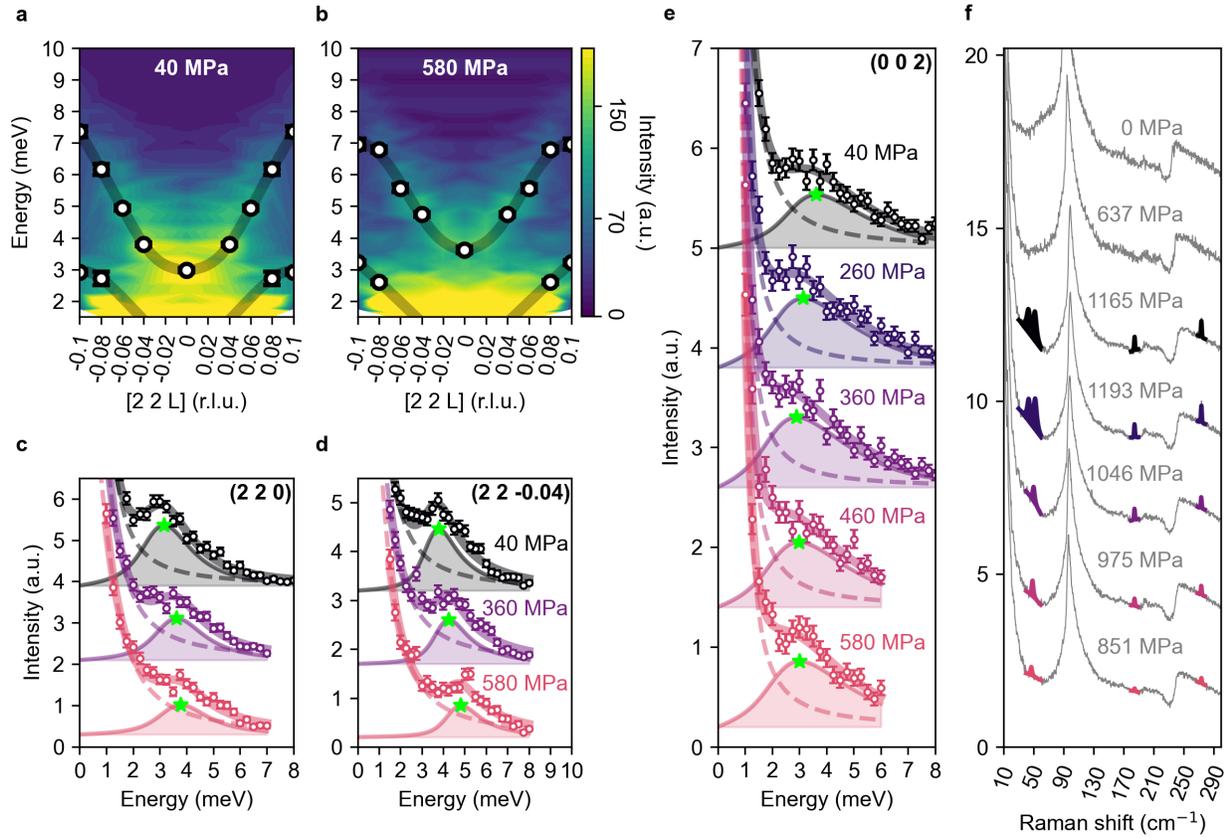

**Figure 2 | Neutron and Raman scattering in strained KTO. (a,b)** Dispersions of the transverse optic (TO) and transverse acoustic (TA) phonons polarized along [110], at two stress levels with stress along [1$\bar{1}$0], measured at 5 K using the HB-3 triple-axis spectrometer (TAS). The TO mode hardens significantly, whereas the TA phonons are only weakly affected by stress. The circles indicate the results of fits to constant-wavevector scans, and error bars represent 1 s.d. The shaded lines are fits to the full dispersion derived in Ref. [50] (see Methods and Figs. S1,2 for details). **(c,d,e)** Energy scans at (2 2 0), (2 2 -0.04) and (0 0 2) at select stresses, with intensities shifted vertically for clarity. The solid lines correspond to fits to the overall instrument response, which is a convolution of the Bragg intensity and TO phonon with the instrument resolution. The dashed lines represent the underlying background, the shaded areas represent the phonon response modeled as a damped harmonic oscillator, and the green stars indicate the corresponding peak positions. A hardening of the phonon along [22$L$] and a slight softening at (002) are observed up to ~600 MPa. **(f)** Raman spectra with stress along [1$\bar{1}$0], measured at 30 K. Two low-energy Raman-active TO modes appear around 800 MPa between 40 cm$^{-1}$ and 60 cm$^{-1}$ (shaded peaks: double-Lorentzian fit), along with two high-energy TO phonons at approximately 185 cm$^{-1}$ and 284 cm$^{-1}$. The broad peak around 90 cm$^{-1}$ is a two-phonon feature known from previous work [27]. Spectra are plotted from top to bottom in order of applied stress and fully reversible upon stress release, demonstrating that the measurements remain within the elastic regime. See Figs. S8-11 for additional data.

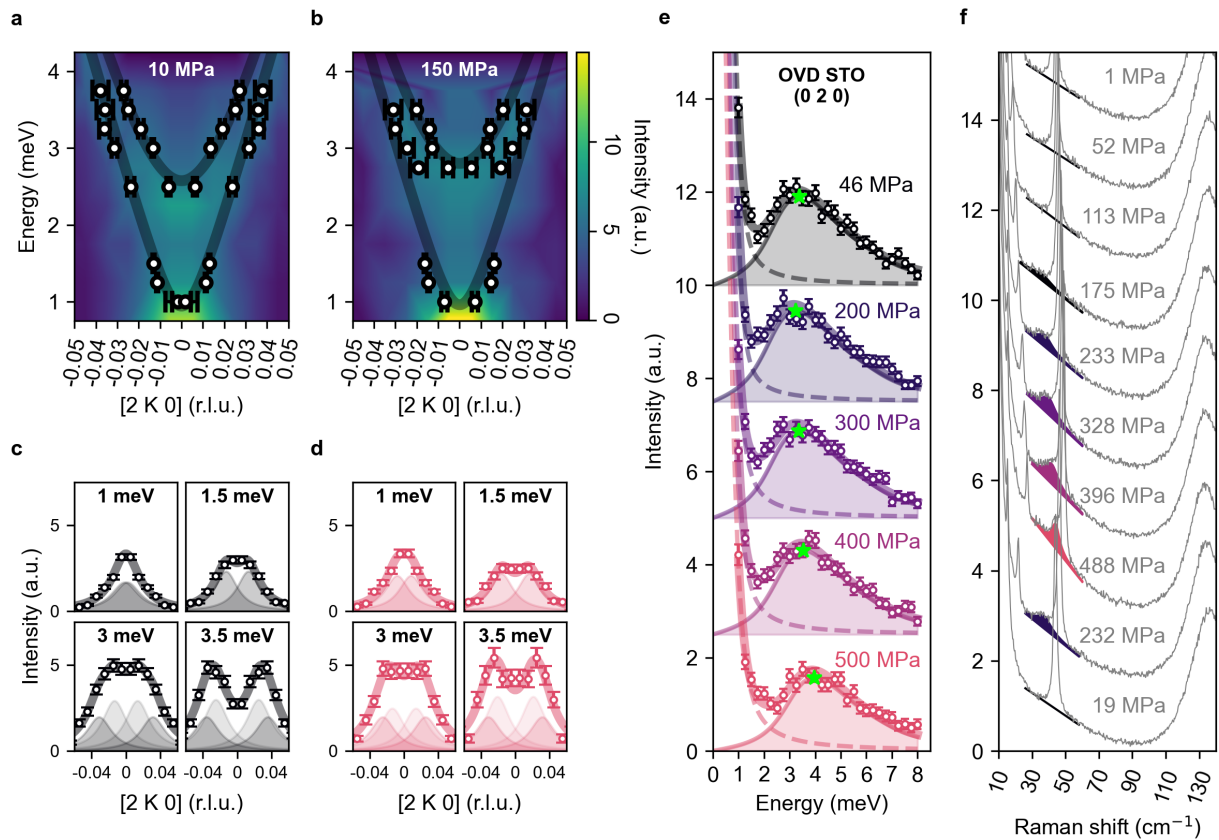

**Figure 3 | Neutron and Raman scattering in strained STO. (a,b)** Dispersions of the $A_{2u}$ and $E_u$ soft TO modes, with stress applied along [001], measured at 2 K in undoped STO using the CNCS time-of-flight (TOF) spectrometer. The circles indicate the fit results, and error bars represent 1 s.d. Close to 2 meV, the fits are unstable (see Methods and Figs. S4-7 for details). Similar to KTO (Fig. 2), the $E_u$ mode hardens slightly, and there is almost no change in the $A_{2u}$ mode close to the critical stress of ~150 MPa. **(c,d)** Momentum cuts of the TOF data in **a,b** along $[2K0]$ at 10 and 150 MPa for select energies. Cuts below 2 meV capture the $A_{2u}$ mode, and cuts above 2 meV capture both the $A_{2u}$ and $E_u$ modes. **(e)** Zone-center TAS energy scans of the [100]-polarized TO phonons in metallic (oxygen vacancy doped, OVD) STO with a carrier density of $n = 1.5 \cdot 10^{19}$ cm$^{-3}$, with stress applied along [001]. Intensities are shifted vertically for clarity. A subtle initial softening is observed up to the critical stress of ~200 MPa, followed by a hardening at higher stresses, as indicated by the green stars (see Fig. S3 for additional data below 200 MPa). **(f)** Raman spectra of OVD-STO ($n = 1.5 \cdot 10^{19}$ cm$^{-3}$) with stress applied along [001], measured at 30 K. The soft TO mode becomes Raman active around 200 MPa (shaded peak: Lorentzian fit), signifying the appearance of long-range inversion-symmetry breaking; the other peaks originate from phonons associated with octahedral rotations in the low-temperature tetragonal phase. A high-energy TO mode at ~170 cm$^{-1}$ is already observed in the unstrained sample, likely due to the presence of dislocations, and increases in intensity above 200 MPa. Spectra are plotted from top to bottom in the order of applied stress and fully reversible upon stress release, demonstrating that the measurements remain within the elastic regime. See Figs. S8-11 for additional data.

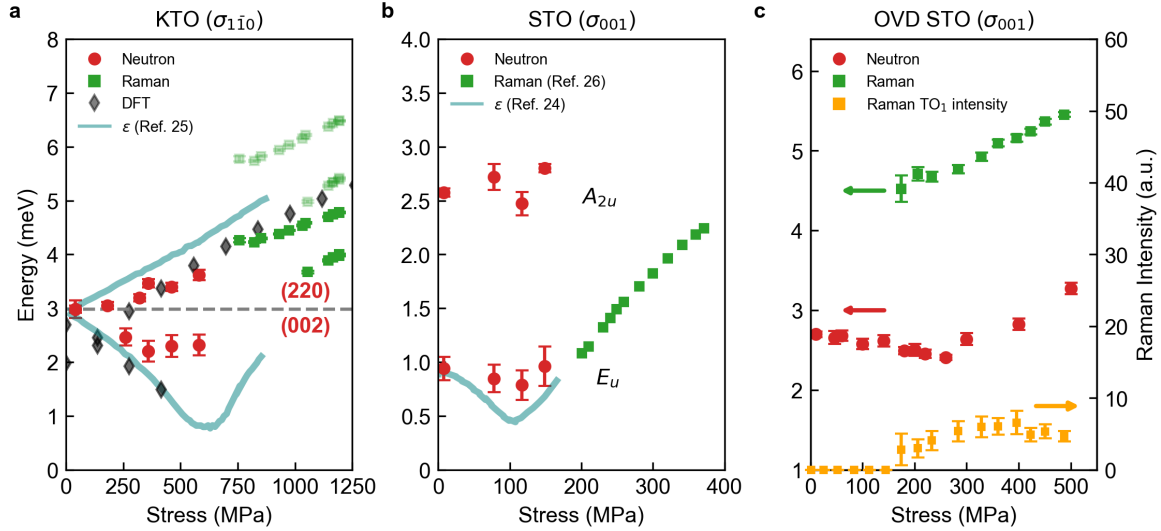

**Fig. 4 | Stress dependence of polar phonons in KTO, STO, and OVD-STO.** (**a**) TO-mode energies in KTO with stress along [1$\bar{1}$0], obtained from neutron scattering (red circles) and Raman scattering (light green squares: data obtained at 30 K; dark green squares: data scaled by temperature dependence measured in [6]), DFT (grey diamonds), and low-frequency dielectric permittivity using the LST relation (solid lines; permittivity data from [25]). The hardening phonon branch shows rather good agreement among neutron scattering, DFT, LST, and the scaled Raman scattering result. In contrast, above ~400 MPa, the softening branch shows a stark discrepancy between the neutron scattering data and the *ab initio* and LST results. (**b**) $A_{2u}$ and $E_u$ TO phonon energies in STO, obtained from neutron scattering (red circles), Raman scattering (green squares; data from [26]) and the LST relation (data from [24]). A breakdown of the LST relation is observed close to the critical stress, in close analogy to KTO. (**c**) Stress-induced changes of the TO mode (red circles, left axis) and integrated Raman intensity from 10 to 100 cm$^{-1}$ (yellow squares, right axis) for OVD-STO. A small initial softening of the phonon polarized perpendicular to the stress direction (red circles) is followed by significant hardening above ~200 MPa, roughly the same stress where the TO phonons become Raman active. In contrast, Raman data for the mode polarized along the stress direction (green squares) show continuous hardening. Similar to KTO, the energy of the latter mode extrapolates to a slightly higher zero-stress value than the neutron scattering results due to higher base temperature of the Raman setup. Unlike for KTO, we did not attempt to scale these data since the temperature dependence of the phonon energies at nonzero stress is not known with sufficient accuracy.

# References

[1] Collignon, C., Lin, X., Rischau, C. W., Fauqué, B. & Behnia, K. Metallicity and superconductivity in doped strontium titanate. *Annu. Rev. Condens. Matter Phys.* **10**, 25 (2019).

[2] Gastiasoro, M. N., Ruhman, J. & Fernandes, R. M. Superconductivity in dilute $SrTiO_3$: a review. *Ann. Phys.* **417**, 168107 (2020).

[3] Weaver, H. E., Dielectric properties of single crystals of $SrTiO_3$ at low temperatures. *J. Phys. Chem. Solids* **11**, 274 (1959).

[4] Fujii, Y. & Sakuda, T. Dielectric and optical properties of $KTaO_3$. *J. Phys. Soc. Jpn.* **41**, 3 (1976).

[5] Yamada, Y. & Shirane, G. Neutron scattering and nature of the soft optical phonon in $SrTiO_3$. *J. Phys. Soc. Jpn.* **26**, 396 (1969).

[6] Shirane, G., Nathans, R. & Minkiewicz, V. J. Temperature dependence of the soft ferroelectric mode in $KTaO_3$. *Phys. Rev.* **157**, 396 (1967).

[7] Müller, K. A. & Burkard, H. $SrTiO_3$: An intrinsic quantum paraelectric below 4 K. *Phys. Rev. B* **19**, 3593 (1979).

[8] Rowley, S. E. et al., Ferroelectric quantum criticality. *Nat. Phys.* **10**, 367 (2014).

[9] Zalar, B. et al. NMR study of disorder in $BaTiO_3$ and $SrTiO_3$. *Phys. Rev. B* **71**, 064107 (2005).

[10] Guzmán-Verri, G. G., Liang, C. H. & Littlewood, P. Lamellar fluctuations melt ferroelectricity. *Phys. Rev. Lett.* **131**, 046801 (2023).

[11] Schooley, J. F., Hosler, W. R. & Cohen, M. L. Superconductivity in semiconducting $SrTiO_3$. *Phys. Rev. Lett.* **12**, 474 (1964).

[12] Lin, X., Zhu, Z., Fauqué, B. & Behnia, K. Fermi surface of the most dilute superconductor. *Phys. Rev. X* **3**, 021002 (2013).

[13] Liu, C. et al. Two-dimensional superconductivity and anisotropic transport at $KTaO_3$ (111) interfaces. *Science* **371**, 716 (2021).

[14] Chen, Z. et al., Electric field control of superconductivity at the $LaAlO_3/KTaO_3$ (111) interface. *Science* **372**, 721 (2021).

[15] Edge, J. M., Kedem, Y., Aschauer, U., Spaldin, N. A. & Balatsky, A. V. Quantum critical origin of the superconducting dome in $SrTiO_3$. *Phys. Rev. Lett.* **115**, 247002 (2015).

[16] Gastiasoro, M.N., Trevisan, T.V. & Fernandes, R.M. Anisotropic superconductivity mediated by ferroelectric fluctuations in cubic systems with spin-orbit coupling. *Phys. Rev. B* **101**, 174501 (2020).

[17] Gastiasoro, M.N., Temperini, M.E., Barone, P. & Lorenzana, J. Theory of superconductivity mediated by Rashba coupling in incipient ferroelectrics. *Phys. Rev. B* **105**, 224503 (2022).

[18] Saha, S. K., Gastiasoro, M.N., Ruhman, J. & Klein, A. Strong coupling theory of superconductivity and ferroelectric quantum criticality in metallic $SrTiO_3$. Preprint at https://arxiv.org/abs/2412.05374 (2024).

[19] Fauqué, B. et al. The polarization fluctuation length scale shaping the superconducting dome of $SrTiO_3$. Preprint at https://arxiv.org/html/2404.04154v1 (2024).

[20] Itoh, M., Wang, R., Narahara, M. & Kyomen, T. Dielectric properties of $SrTi^{18}O_3$. *Ferroelectrics* **285**, 3 (2003).

[21] Rischau, C. W. et al. A ferroelectric quantum phase transition inside the superconducting dome of $Sr_{1-x}Ca_xTiO_{3-\delta}$. *Nat. Phys.* **13**, 643 (2017).

[22] Tomioka, Y., Shirakawa, N. & Inoue, I. H. Superconductivity enhancement in polar metal regions of $Sr_{0.95}Ba_{0.05}TiO_3$ and $Sr_{0.985}Ca_{0.015}TiO_3$ revealed by systematic Nb doping. *NPJ Quant. Mater.* **7**, 111 (2022).

[23] Axelsson, A. K., Pan, Y., Valant, M., Vilarinho, P.M. & Alford, N.M. Polar fluctuations in Mn substituted $KTaO_3$ ceramics. *J. Appl. Phys.* **108**, 064109 (2010).

[24] Burke, W. J. & Pressley, R. J. Stress induced ferroelectricity in $SrTiO_3$. *Solid State Commun.* **9**, 191 (1971).

[25] Uwe, H. & Sakudo, T. Electrostriction and stress-induced ferroelectricity in $KTaO_3$. *J. Phys. Soc. Jpn.* **38**, 1 (1975).

[26] Uwe, H. & Sakudo, T. Stress-induced ferroelectricity and soft phonon modes in $SrTiO_3$. *Phys. Rev. B* **13**, 271–286 (1976).

[27] Uwe. H. & Sakudo, T. Raman-scattering study of stress-induced ferroelectricity in $KTaO_3$. *Phys. Rev. B* **15**, 1 (1977).

[28] Haeni, J.-H. et al., Room-temperature ferroelectricity in strained $SrTiO_3$. *Nature* **430**, 758 (2004).

[29] Ahadi, K. et al. Enhancing superconductivity in $SrTiO_3$ films with strain. *Sci. Adv.* **5**, eaaw0120 (2019).

[30] Salmani-Rezaie, S., Kim, H., Ahadi, K. & Stemmer, S. Lattice relaxations around individual dopant atoms in $SrTiO_3$. *Phys. Rev. Materials* **3**, 114404 (2019).

[31] Hameed, S. et al., Enhanced superconductivity and ferroelectric quantum criticality in plastically deformed strontium titanate. *Nat. Mater.* **21**, 54 (2022).

[32] Khayr, I. et al., Structural properties of plastically deformed $SrTiO_3$ and $KTaO_3$. *Phys. Rev. Materials* **8**, 124404 (2024).

[33] Herrera, C., Cerbin, J., Jayakody, A., Dunnett, K., Balatsky, A. V. & Sochnikov, I. Strain-engineered interaction of quantum polar and superconducting phases. *Phys. Rev. Materials* **3**, 124801 (2019).

[34] Li, J. et al. The classical-to-quantum crossover in the strain-induced ferroelectric transition in $SrTiO_3$ membranes. *Nat. Commun.* **16**, 4445 (2025).

[35] Dunnett, K., Narayan, A., Spaldin N. A. & Balatsky, A. V. Strain and ferroelectric soft-mode induced superconductivity in strontium titanate. *Phys. Rev. B* **97**, 144506 (2018).

[36] Tomioka, Y., Shirakawa, N., Shibuya, K. & Inoue, I. H. Enhanced superconductivity close to a non-magnetic quantum critical point in electron-doped strontium titanate. *Nat. Commun.* **10**, 738 (2019).

[37] Klein, A., Kozii, V., Ruhman, J. & Fernandes, R. M. Theory of criticality for quantum ferroelectric metals. *Phys. Rev. B* **107**, 165110 (2023).

[38] Salmani-Rezaie, S., Ahadi, K., Strickland, W. M. & Stemmer, S. Order-disorder ferroelectric transition of strained $SrTiO_3$. *Phys. Rev. Lett.* **125**, 087601 (2020).

[39] Chu, J.-H. et al. In-plane resistivity anisotropy in an underdoped iron arsenide superconductor. *Science* **329**, 824–826 (2010).

[40] Pfeifer, E. R. & Schooley, J. F., Effect of stress on the superconductive transition temperature of strontium titanate. *Phys. Rev. Lett.* **19**, 783 (1967).

[41] Steppke, A. et al. Strong peak in $T_c$ of $Sr_2RuO_4$ under uniaxial pressure. *Science* **335**, 9398 (2017).

[42] Najev. A. et al. Uniaxial strain control of bulk ferromagnetism in rare-earth titanates. *Phys. Rev. Lett.* **128**, 167201 (2022).

[43] Lieberich, F., Saito, Y., Agarmani, Y., Sasaki, T., Yoneyama, N., Winter, S. M., Lang, M. & Gati, E. Probing and tuning geometric frustration in an organic quantum magnet via elastocaloric measurements under strain. *Sci. Adv.* **11**, eadz0699 (2025).

[44] Yamanaka, A., Kataoka, M., Inaba, Y., Inoue, K., Hehlen, B. & Courtens, E. Evidence for competing orderings in strontium titanate from hyper-Raman scattering spectroscopy. *Europhys. Lett.* **50**, 688 (2000).

[45] Van Mechelen, J.L.M., Van der Marel, D., Grimaldi, C., Kuzmenko, A.B., Armitage, N.P., Reyren, N., Hagemann, H. & Mazin, I.I. Electron-phonon interaction and charge carrier mass enhancement in $SrTiO_3$. *Phys. Rev. Lett.* **100**, 226403 (2008).

[46] Shirane, G., Shapiro, S.M., Tranquada, J.M. Neutron scattering with a triple-axis spectrometer. *Cambridge University Press*.

[47] Lyddane, R. H., Sachs, R. G. & Teller, E. On the polar vibrations of alkali halides. *Phys. Rev.* **59**, 673 (1941).

[48] Larkin, A. I. & Pikin, S. A. Phase transitions of the first order but nearly of the second. *J. Exp. Theor. Phys.* **29**, 891 (1969).

[49] He, X., Bansal, D., Winn, B., Chi, S., Boatner, L & Delaire, O. Anharmonic eigenvectors and acoustic phonon disappearance in quantum paraelectric $SrTiO_3$. *Phys. Rev. Lett.* **124**, 145901 (2020).

[50] Axe, J. D., Harada, J. & Shirane, G. Anomalous acoustic dispersion in centrosymmetric crystals with soft optic phonons. *Phys. Rev. B* **1**, 1227 (1970).

[51] Fauqué, B. et al. Mesoscopic fluctuating domains in strontium titanate. *Phys. Rev. B* **106**, L140301 (2022).

[52] Herrera, C. et al. Strain-engineered interaction of quantum polar and superconducting phases. *Phys. Rev. Mater.* **3**, 124801 (2019).

[53] Yamada, Y. & Shirane, G. Study of critical fluctuations in BaTiO$_3$ by neutron scattering. *Phys. Rev.* **177**, 2 (1969).

[54] Wang, X. et al. Multiferroicity in plastically deformed SrTiO$_3$. *Nat. Commun.* **15**, 7442 (2024).

[55] Zhong, R. D., Schneeloch, J.A., Liu, T.S., Camino, F.E., Tranquada, J.M. & Gu, G.D. Superconductivity induced by In substitution into topological crystalline insulator Pb$_{0.5}$Sn$_{0.5}$Te. *Phys. Rev. B* **90**, 020505(R) (2014).

[56] Narayan, A., Cano, A., Balatsky, A.V. & Spaldin, N.A. Multiferroic quantum criticality. *Nat. Mater.* **18**, 223 (2019).

[57] Lin, X. et al. Critical doping for the onset of a two-band superconducting ground state in SrTiO$_{3-\delta}$. *Phys. Rev. Lett.* **112**, 207002 (2014).

[58] Ehlers, G., Podlesnyak, A.A., Niedziela, J.L., Iverson, E.B. & Sokol, P.E. The new cold neutron chopper spectrometer at the Spallation Neutron Source: Design and performance. *Rev. Sci. Instrum.* **82**, 085108 (2011).

[59] Zheludev, A. Reslib 3.4 software. *Oak Ridge National Laboratory* (2007).

[60] Cochran, W. & Cowley, R.A. Dielectric constants and lattice vibrations. *J. Phys. Chem. Solids* **23**, 447 (1962).

[61] Perdew, J. P., Burke, K., Ernzerhof, M., Generalized Gradient Approximation. *Phys. Rev. Lett.* **77**, 3865 (1996).

[62] Blochl, P.E. Projector augmented-wave method. *Phys. Rev. B* **50**, 17953 (1994).

[63] Kresse, D. J. G. From ultrasoft pseudopotentials to the project augmented-wave method. *Phys. Rev. B* **59**, 1758 (1999).

# Supplementary Information for:

# Uniaxial strain tuning of polar lattice vibrations in KTaO$_3$ and SrTiO$_3$

I. Khayr[1*], N. Somun[2*], S. Hameed[3*], Z. Van Fossan[4], X. He[1], R. Spieker[1], S. Chi[5], E. Clements[5], D. Pajerowski[5], M. Minola[3], B. Keimer[3], T. Birol[4], D. Pelc[1,2†], and M. Greven[1†]

This document contains nine sections with additional information pertaining to data analysis procedures as well as additional triple-axis neutron spectroscopy, time-of-flight neutron scattering, and Raman scattering data, along with supporting tables and figures.

Section I: fit parameters used to analyze triple-axis neutron data

Section II: deconvolution procedure used to determine phonon energies and dispersions from constant-wavevector (energy) triple-axis neutron scans

Section III: additional constant-wavevector measurements of KTaO$_3$ (KTO) and SrTiO$_3$ (STO)

Section IV: raw data and fits to time-of-flight neutron scattering data

Section V: additional Raman data for KTO and STO for [$1\bar{1}0$] and [001] stress, respectively, that are not included in Figs. 2 and 3 of the main text

Section VI: procedure for stress calibration in *in situ* stress Raman scattering measurements

Section VII: Raman scattering data for a STO sample that was strained beyond the elastic regime

Section VIII: Raman scattering data for KTO with stress along [001]

Section IX: additional density functional theory (DFT) calculations for KTO with stress along both [$1\bar{1}0$] and [001]

# I. Fit procedure for constant-wavevector scans

Constant-wavevector energy scans were fit to one gaussian (Bragg peak) and one or two DHOs (TO/TA phonons), with a total of five (or eight, if both TO and TA modes are observed) free parameters. In order to determine the stress dependence of the phonon parameters, we relaxed all parameters at the lowest stress, then fixed various parameters (except the phonon energy) to assess the behavior of the phonon energy stress dependence and the influence of other parameters (e.g., phonon amplitude and linewidth). We tried four different configurations, as shown in Fig. S1: relaxing all parameters (R), fixing $\Gamma$ (FG), fixing the Bragg intensity (FB), fixing both the Bragg intensity and $\Gamma$ (FBG), and fixing $\Gamma$ and the phonon intensity (FGA). Figure S1a reveals that, in all cases, the (002) phonon softens to some degree, regardless of the parameter settings, indicative of a robust effect. Additionally, there are no significant changes in the linewidths and amplitudes for all four configurations (Fig. S1b,c). In the case of the (220) phonon (Fig. S1d-f), the FB and FBG configurations show a peak in the phonon energy around 380 MPa, but based on our *ab initio* calculation and previous dielectric measurements (Fig. 4a), we expect this phonon to monotonically harden with stress. We therefore report the FG result, since it matches the expected behavior at both (220) and (002). In OVD STO, the stress dependence of the phonon energy (Fig. 3e) is essentially the same independent of which parameters are fixed or relaxed. Results are shown in Fig. S1g-i.

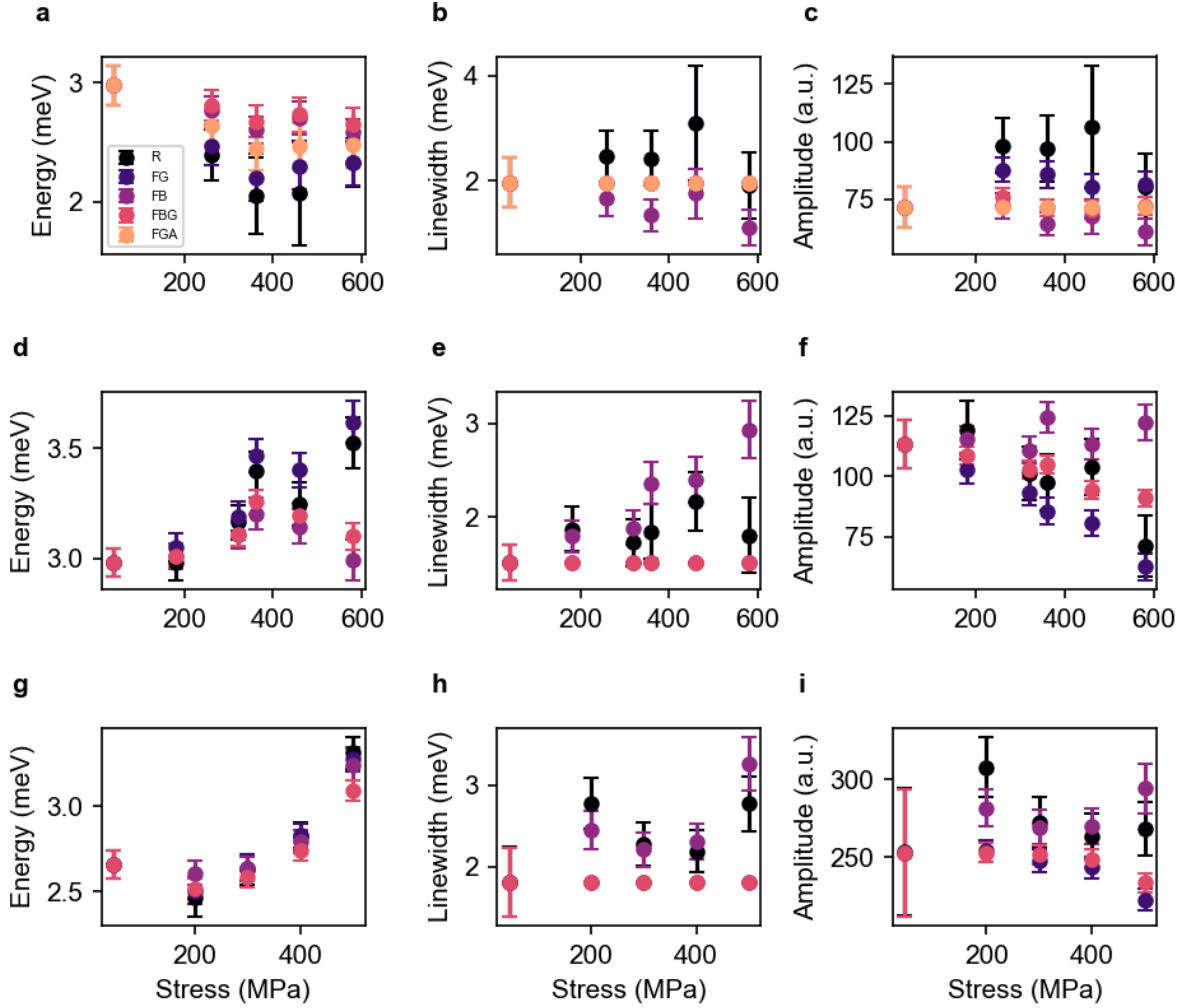

**Fig. S1:** Fit results for various parameter settings: Relaxing all parameters (R), fixing Γ (FG), fixing the Bragg intensity (FB), fixing both the Bragg intensity and Γ (FBG), and fixing Γ and the phonon intensity (FGA). (**a-c**) Stress dependence of phonon energy, linewidth, and amplitude, respectively, for the KTO (002) phonon. (**d-f**) Stress dependence of phonon energy, linewidth, and amplitude, respectively, for the KTO (220) phonon. (**g-h**) Stress dependence of phonon energy, linewidth, and amplitude for the STO (020) phonon.

## II. Determination of phonon energies and dispersions from triple-axis spectroscopy

The MATLAB library ResLib was used to perform deconvolution of constant-wavevector data obtained on the HB-3 triple-axis spectrometer at the High Flux Isotope Reactor, Oak Ridge National Laboratory. The deconvolution requires an input model to characterize the features in the energy spectra. In this case, our model consists of the Bragg peak and the transverse-optic phonon contribution in both KTO and STO. The Bragg peak is taken to be a Lorentzian centered at zero energy and with variable width, and the phonon is modeled by a damped harmonic oscillator (DHO) function, scaled by the thermal occupation factor $n$:

$$I(q,\omega) = (n+1)\frac{A\omega}{\left(\omega^2 - \omega_1^2(q)\right)^2 + \Gamma^2\omega^2} \quad (1)$$

where $A$, $\omega_1(q)$, and $\Gamma$ are free parameters that represent the amplitude, frequency, and linewidth of the phonon, respectively. The phonon dispersion observed in time-of-flight measurements of STO (Fig. 3a,b) is steep compared to the tilt of the HB-3 resolution ellipsoid, and the convolution process thus includes intensity from the phonon branch at higher $q$. It was therefore necessary to include the $q$-dependence of $\omega_1$ in our model. For small $q$, we assume the phonon dispersion

$$\omega_1^2 = \omega_{TO}^2 + v_{TO}^2 q^2 \quad (2)$$

where $\omega_{TO}$ is the characteristic phonon energy and $v_{TO}$ is the phonon velocity parameter. In order to determine $v_{TO}$ in KTO and STO, we initially performed fits of the constant-wavevector spectra along [22$L$] and [$H$20], respectively (Fig. S2 a,b and Fig. S3a) under the assumption that $\omega_1$ is $q$-independent, i.e., $\omega_1 = \omega_{TO}$. We then extracted the velocity parameter by fitting the phonon energies at the corresponding $q$ to (2). Once we had obtained $v_{TO}$, we repeated the fit for the zone-center spectra using (2). The values of $v_{TO}^2$ for KTO and STO are tabulated in Table I. For the (002) phonon in KTO, we scaled the value of $v_{TO}^2$ so that the extracted phonon energy from the fit matched the phonon energy at (220). The discrepancy in $v_{TO}^2$ values can be attributed to the known anisotropic behavior of the phonon dispersion in KTO [S1].

| **Material (Phonon)** | $v_{TO}^2$ (meV/r.l.u.)² |
|---|---|
| KTO (220) | $4253 \pm 125$ |
| KTO (002) | $6460 \pm 125$ |
| STO (020) | $7420 \pm 250$ |

**Table. S1:** Values of $v_{TO}^2$ used in the input model for the extraction of the TO phonon energies in KTO and STO.

When refitting the energy spectra measured along [22$L$] using the $v_{TO}^2$ values in Table I, we used a linear approximation for the dispersion around a given value of $q$. A Taylor series approximation of Eq. 2 results in

$$\omega_1 = \frac{1}{\omega_{TO}}\left(\omega_{TO}^2 + A(L^2 - qL)\right). \tag{3}$$

This expression was used in our model for $L$ = -0.04 to -0.10 r.l.u. (see Fig. S2 c-f). $A$ is equivalent to the square of the phonon velocity $v_{TO}$.

The lines in Fig. 2a,b represent fits that use the full dispersion relation and that take acoustic-optic coupling into account, as derived in [S2]. The following determinant was solved to extract the matrix elements, in this case, $f_{11}$, $f_{12}$, and $f_{22}$:

$$\begin{vmatrix} \omega_o^2 + f_{11} - \omega^2(q) & f_{12}(q) \\ f_{12}(q) & f_{22}(q) - \omega^2(q) \end{vmatrix} = 0 \tag{4}$$

where $f_{ij} = F_{ij}^{(2)}q^2 + F_{ij}^{(4)}q^4 + \cdots$. Values of $f_{ij}$ used for the dispersion in Fig. 2a,b are tabulated in Table II.

|  | $F_{11}^{(2)}$ | $F_{11}^{(4)}$ | $F_{22}^{(2)}$ | $F_{22}^{(4)}$ | $F_{12}^{(2)}$ | $F_{12}^{(4)}$ | $\omega_{TO}$ |
|---|---|---|---|---|---|---|---|
| **40 MPa** | 2193 | 97926 | 5468 | -293883 | 2083 | 10653 | 2.98 |
| **580 MPa** | 4174 | -205025 | 5747 | -251666 | 3786 | 223734 | 3.63 |

**Table. S2:** Values of $f_{ij}$ (meV/r.l.u.)$^2$ and $\omega_{TO}$ (meV) used for the dispersion in Fig. 2a,b at 40 and 580 MPa.

## III. Additional triple-axis data

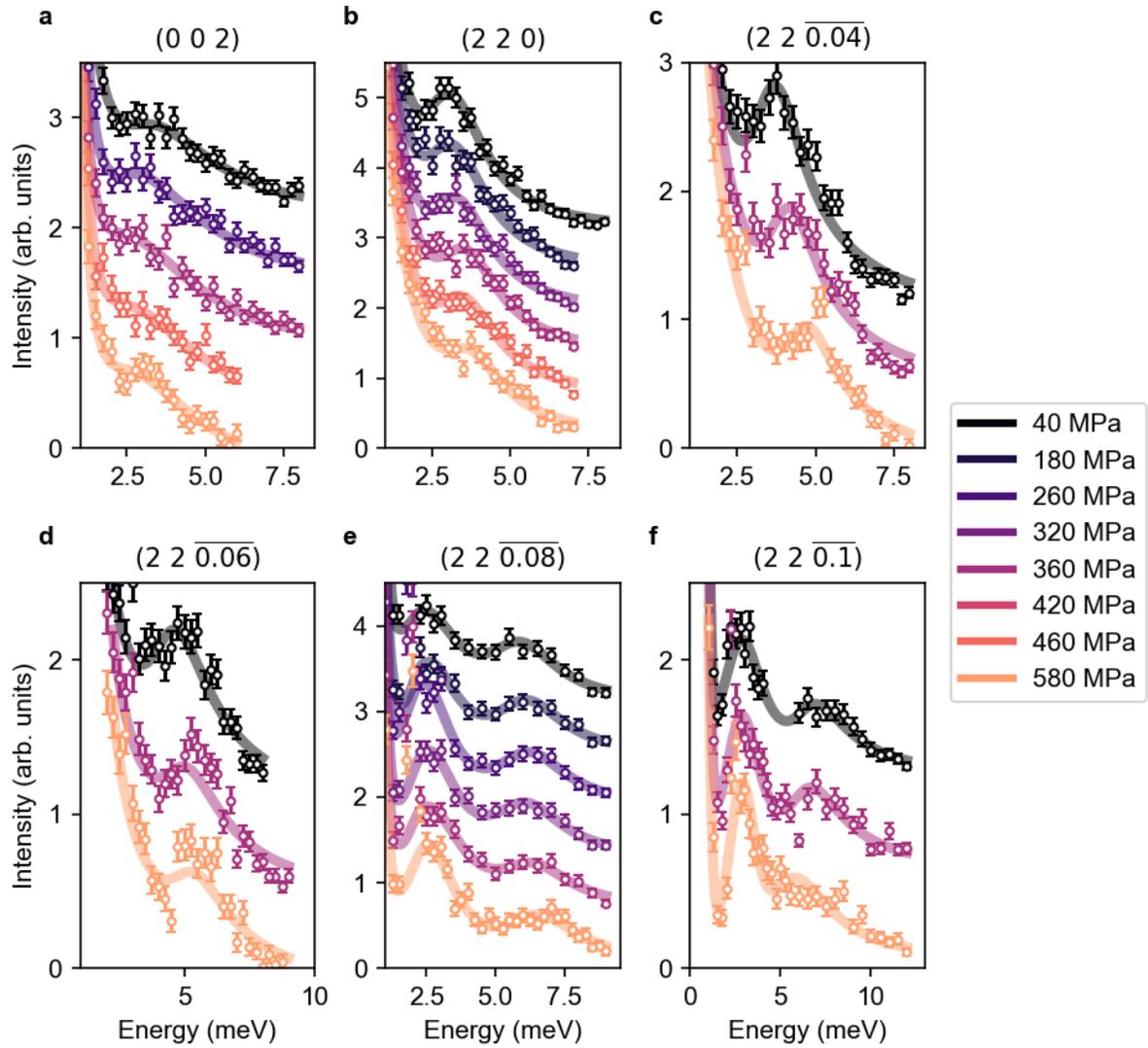

**Fig. S2:** Inelastic neutron scattering data and corresponding fit results for strained KTO, with lighter colors corresponding to higher stresses, as indicated. (**a**) Energy dependence at (002). A slight initial softening of the TO phonon mode is observed. (**b-f**) Energy dependence at various $L$ values [22$L$]. All results are shifted vertically for clarity.

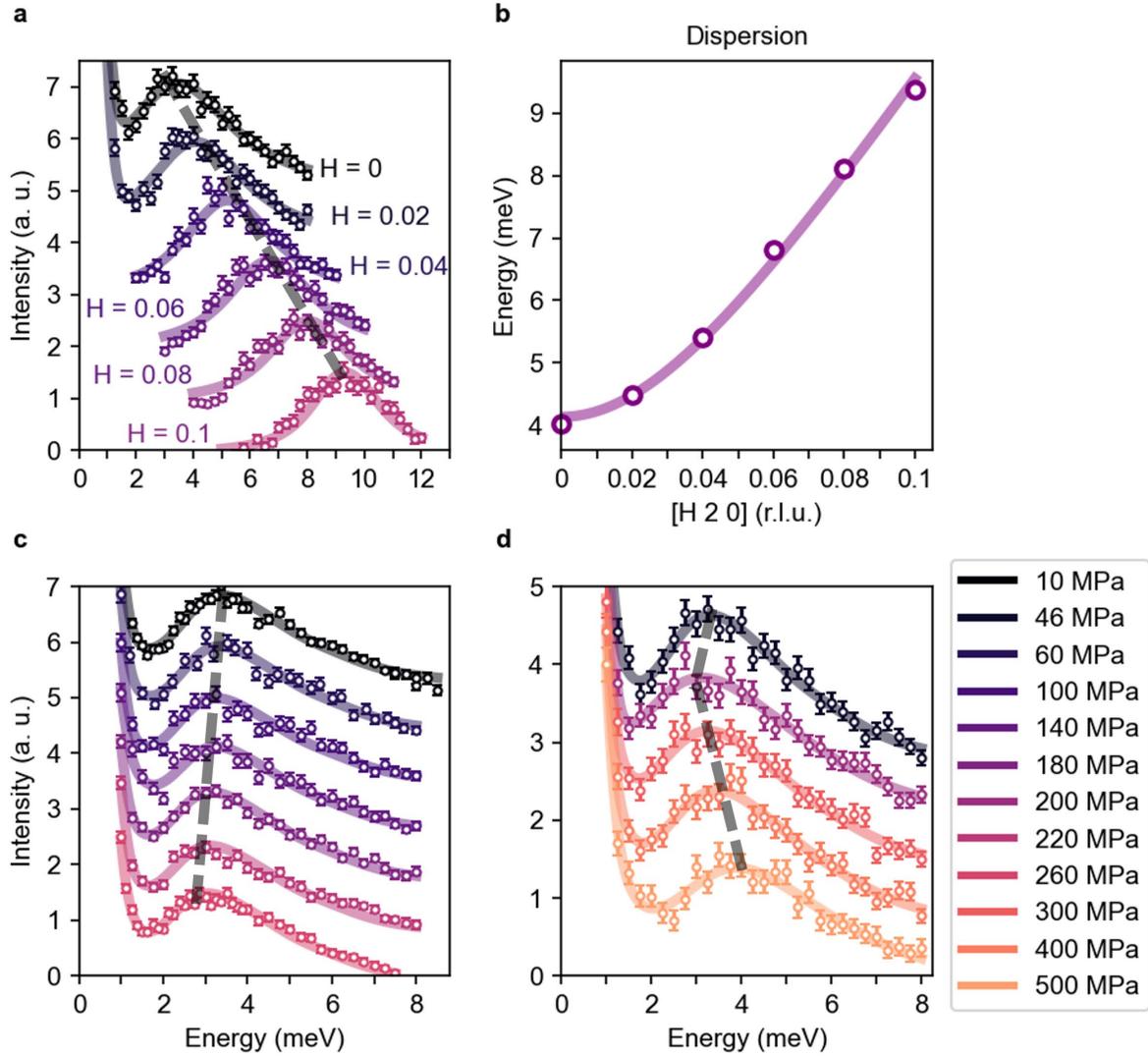

**Fig. S3:** Energy dependence of neutron scattering intensity and corresponding fit results in strained STO-OVD, with lighter colors corresponding to higher stresses. (**a**) Energy dependence along [*H*20] for different *H* values. (**b**) Extracted phonon energies from spectra in **a**. From this, we can determine the slope of the dispersion, which is included in our model when performing a deconvolution. (**c**) Energy dependence for OVD-STO with maximum achieved stress of 260 MPa. (**d**) Energy dependence for a second OVD-STO sample, with maximum achieved stress of 500 MPa. Fits in **c** and **d** use the dispersion slope determined from **b** to account for the resolution function capturing the phonon away from the Brillouin zone center.

## IV. Additional time-of-flight data

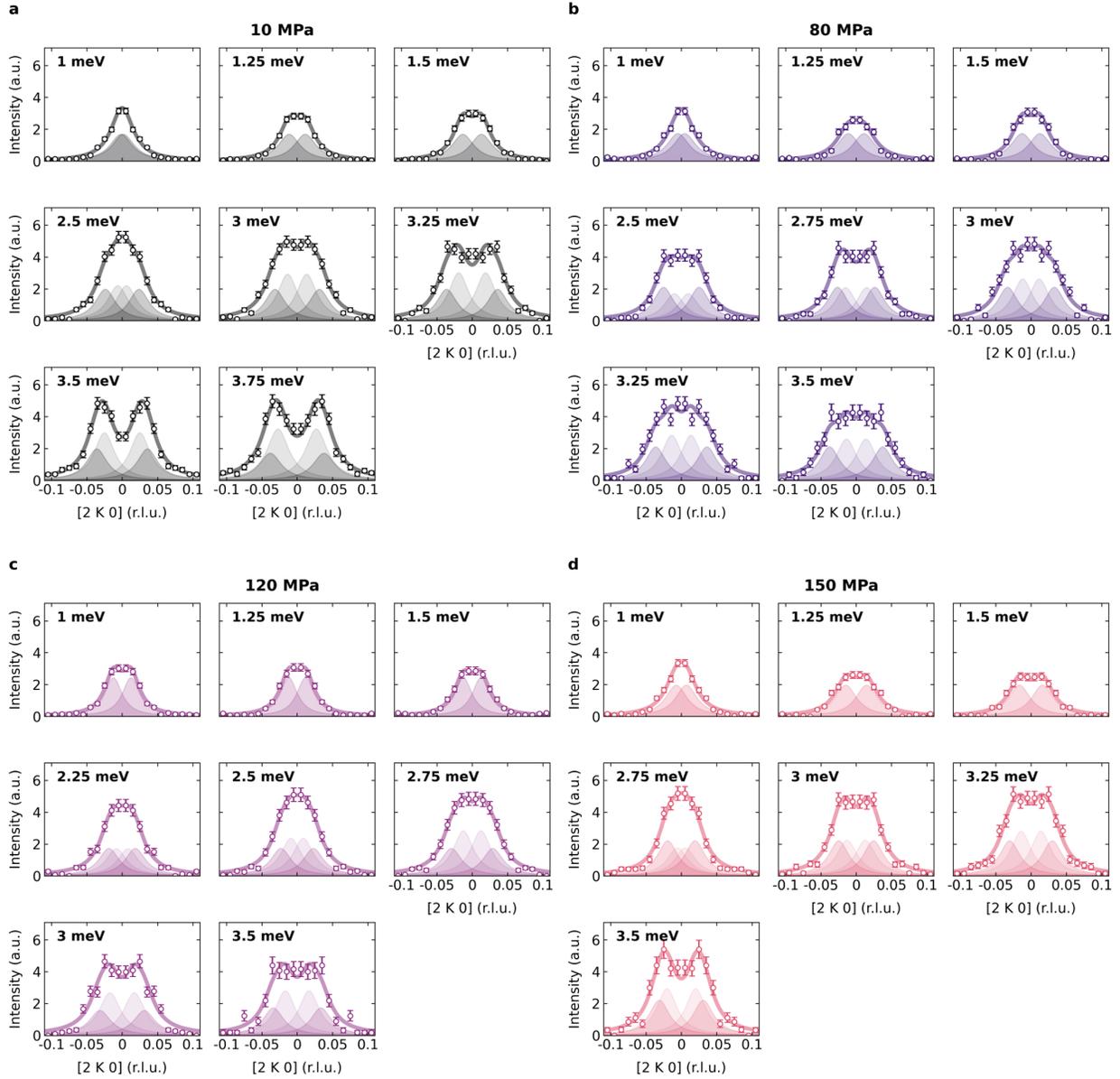

**Fig. S4:** Inelastic neutron scattering data collected at CNCS with an incident energy of 8.5 meV along with corresponding fit results for strained, undoped STO. **(a-d)** Constant energy cuts along [2$K$0] at 10, 80, 120, and 150 MPa with energy transfers ranging from 1 to 3.5 meV. Below 2 meV, two DHOs are fit to the spectra, representing the $A_{2u}$ mode. Above 2 meV, the spectra are fit to both the $A_{2u}$ and $E_u$ modes.

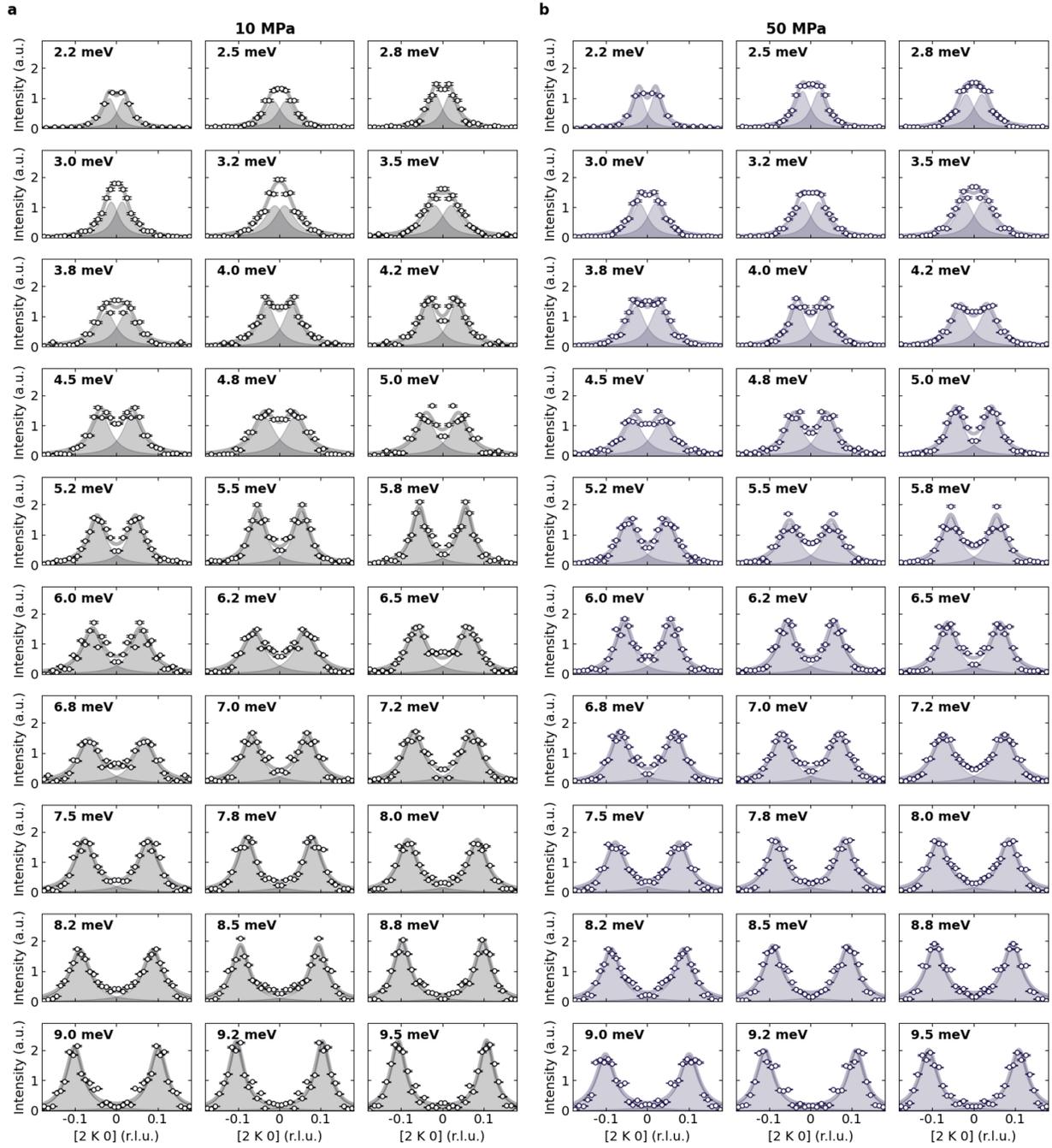

**Fig. S5:** Inelastic neutron scattering data collected at CNCS with an incident energy of 12 meV along with corresponding fit results for strained, undoped STO. Constant energy cuts along [2$K$0] at **(a)** 10 and **(b)** 50 MPa with energy transfers ranging from 2.2 to 9.5 meV.

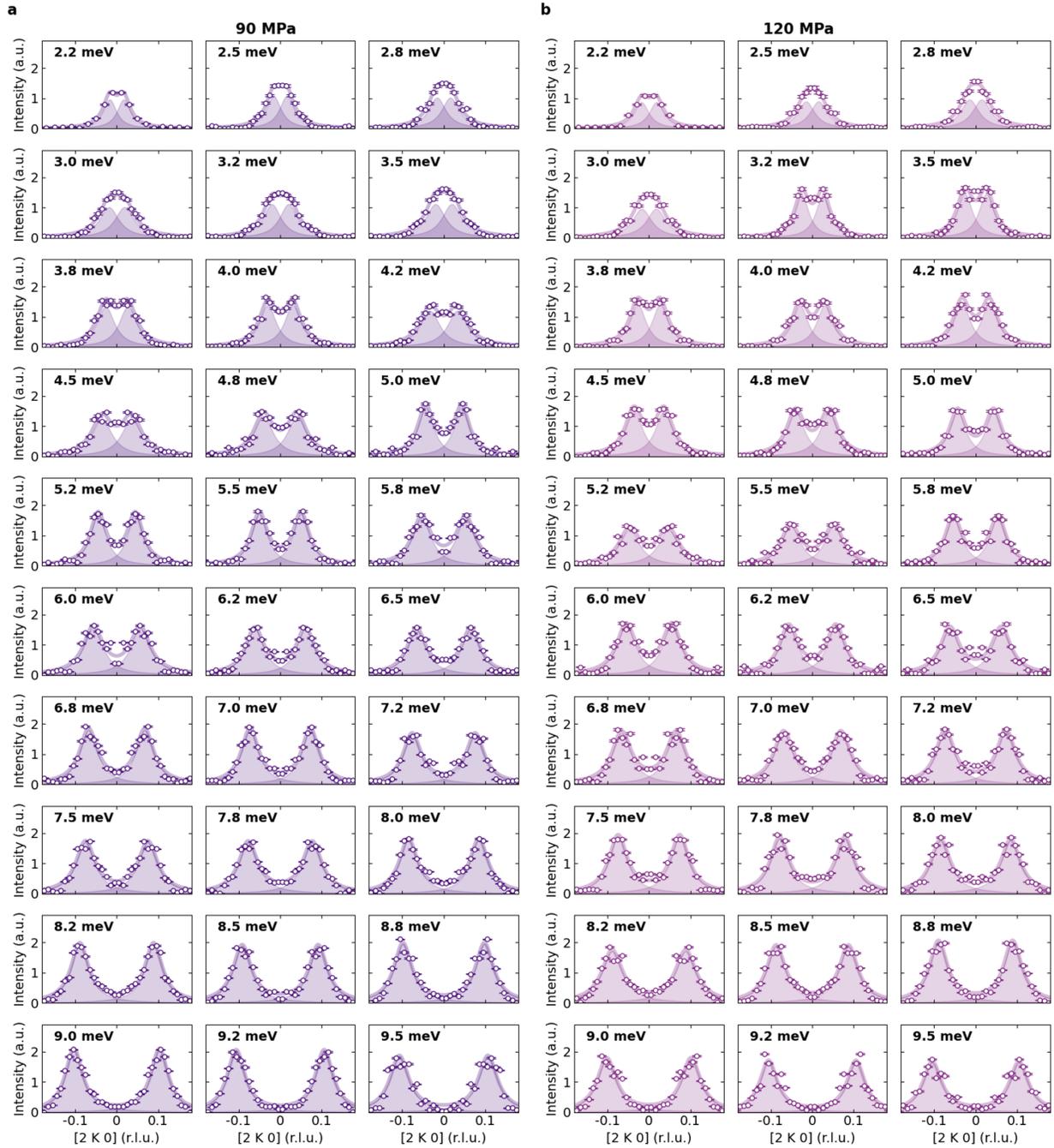

**Fig. S6:** Inelastic neutron scattering data collected at CNCS with an incident energy of 12 meV along with corresponding fit results for strained, undoped STO. Constant energy cuts along [2$K$0] at **(a)** 90 and **(b)** 120 MPa with energy transfers ranging from 2.2 to 9.5 meV.

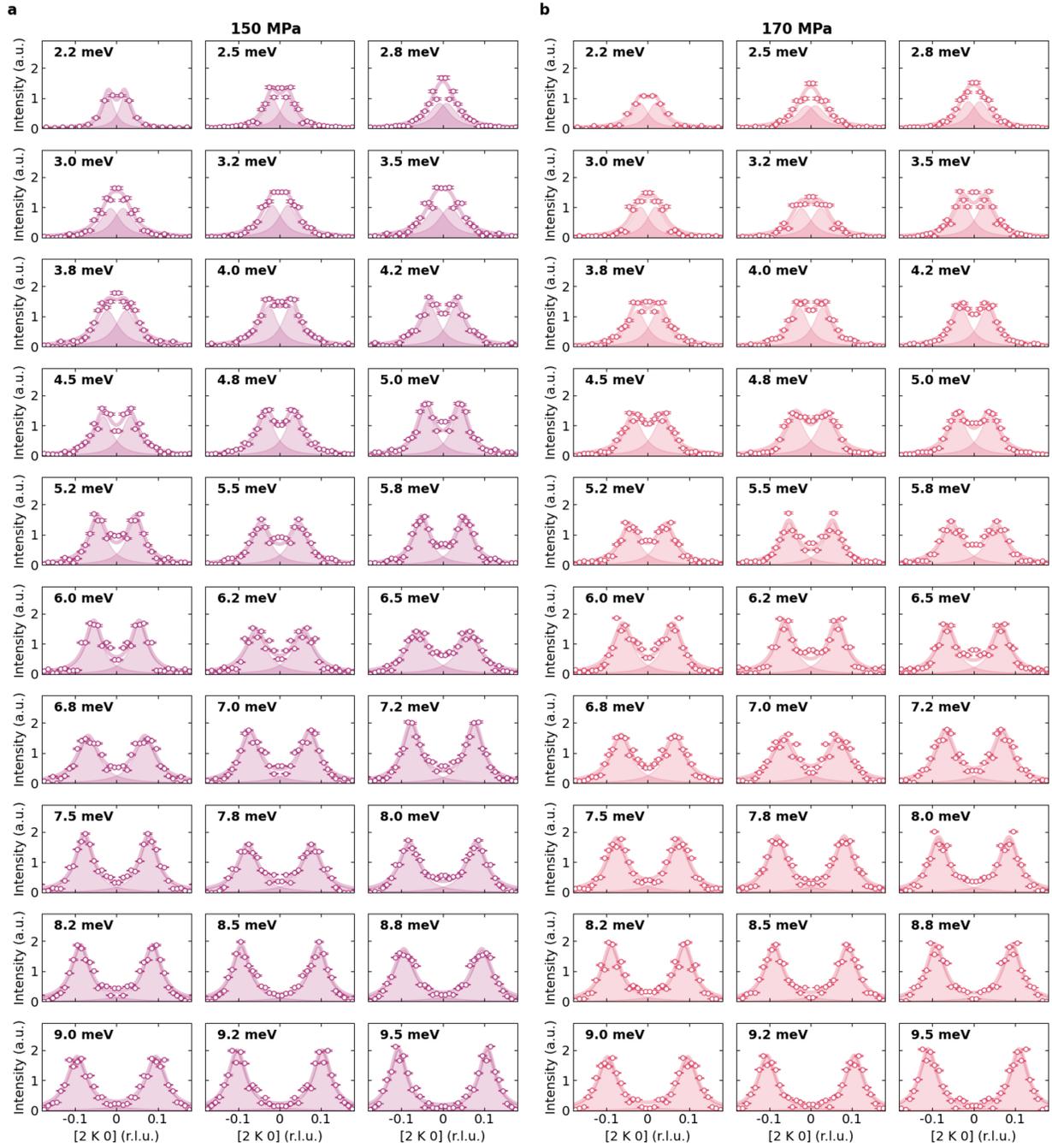

**Fig. S7:** Inelastic neutron scattering data collected at CNCS with an incident energy of 12 meV along with corresponding fit results for strained, undoped STO. Constant energy cuts along [2$K$0] at **(a)** 150 and **(b)** 170 MPa with energy transfers ranging from 2.2 to 9.5 meV.

# V. Additional Raman measurements for $[1\bar{1}0]$ stress on KTO and $[001]$ stress on STO

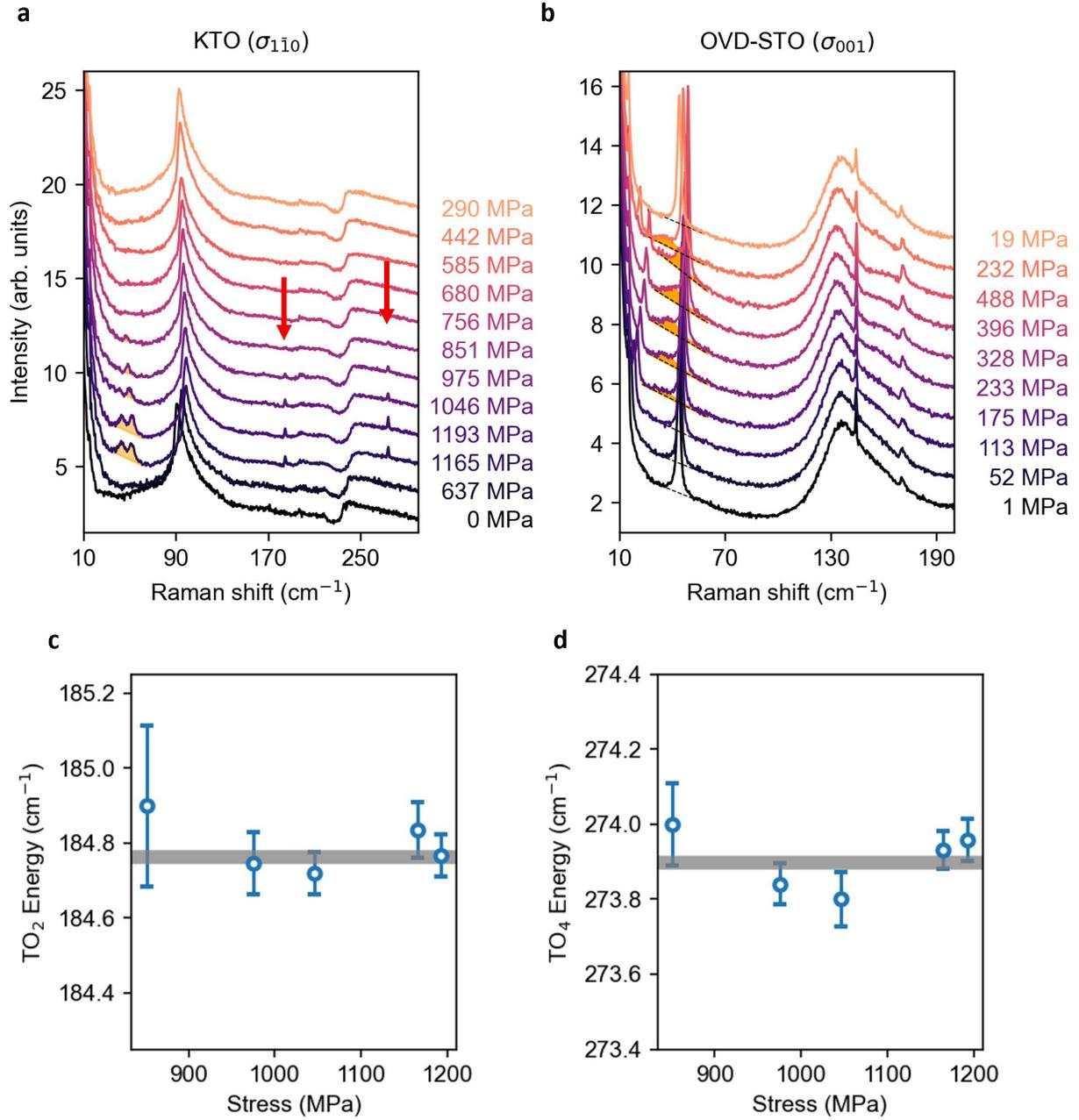

**Fig. S8: (a,b)** Additional data for strained KTO and STO not included in Figs. 2 and 3 of the main text. Spectra are plotted from bottom to top in order of applied stress. **(c,d)** Stress dependence of the $TO_2$ and $TO_4$ modes in strained KTO (see red arrows in **a,b**). There is essentially no change in the energies of both of these modes at elevated stresses.

## VI. Stress calibration for Raman measurements

In the piezo-electric strain device that is used for Raman measurements, the sample stress/strain is controlled by applying different voltages to the built-in piezoelectric stacks. This results in capacitance changes of a displacement sensor, which is based on a parallel-plate capacitor. Using the inverse relation between the capacitance value and the displacement of a parallel-plate capacitor ($C = \frac{A\epsilon_0}{d}$, where $C$ is the capacitance, $\epsilon_0$ is the vacuum permittivity, $A$ is the area of the parallel plates, and $d$ is the distance between the plates), one obtains displacement changes for each strain step. However, the precise strain transmitted to the sample can depend on the sample width and thickness, as well as the amount of epoxy used to glue the sample to the carrier. Therefore, an independent calibration of the stress/strain level is required.

For the case of [001] strain in OVD-STO, the $A_{1g}$ phonon mode observed at ~ 44 cm$^{-1}$ exhibits a strong dependence on stress. According to prior work, the squared frequency of this phonon mode varies with stress $P$ as $\omega^2 = \omega_0^2 + 1.002\,P$, where $\omega_0$ is the frequency at zero stress [S3]. To extract the peak position of this phonon mode, a Lorentzian function is used, as illustrated in Fig. S9(a). The shift in peak position relative to the zero-stress reference is then used to determine the reported stress levels.

Similarly, for the case of [001] strain in KTO, the second-order TA phonon peak at ~ 90 cm$^{-1}$ has been reported in the literature to vary linearly with stress, with a slope of ~ 0.0168 cm$^{-1}$/MPa [S4]. Due to its asymmetric lineshape, a Fano lineshape function is used to extract the peak position, as shown in Fig. S9(b). The extracted peak shift with respect to the zero-stress reference is then used to determine the reported stress levels.

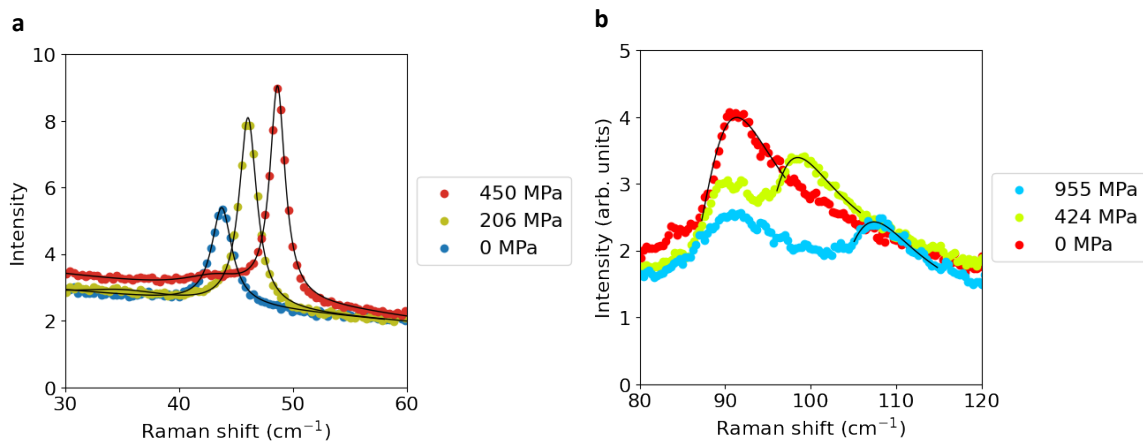

**Fig. S9**: (a) Raman spectra of the $A_{1g}$ phonon at ~ 44 cm$^{-1}$ measured at different stresses along [001] in OVD-STO. The solid lines are fits to the data. For zero stress, a single Lorentzian is used, whereas a double-Lorentzian is used for 206 MPa and 450 MPa to account for the soft TO1 mode that appears above 200 MPa. (b) Raman spectra of the second-order TA phonon peak at ~ 90 cm$^{-1}$ measured at different stresses along [001] in KTO. The solid lines are fits of the higher-energy peak using a Fano asymmetric lineshape. Note that the Fano lineshape is used purely because it provides a good fit and does not carry any physical significance. For both cases, the stress values are obtained from the shift of the phonon peak compared to the peak position for the unstrained "0 MPa" spectrum. All spectra were obtained at 30 K.

For the case of [1$\bar{1}$0] strain on KTO, no calibration of the stress/strain dependence of various phonon modes exist in the literature. Therefore, we carried out independent x-ray diffraction (XRD) measurements on the same sample that was used for the Raman measurements.

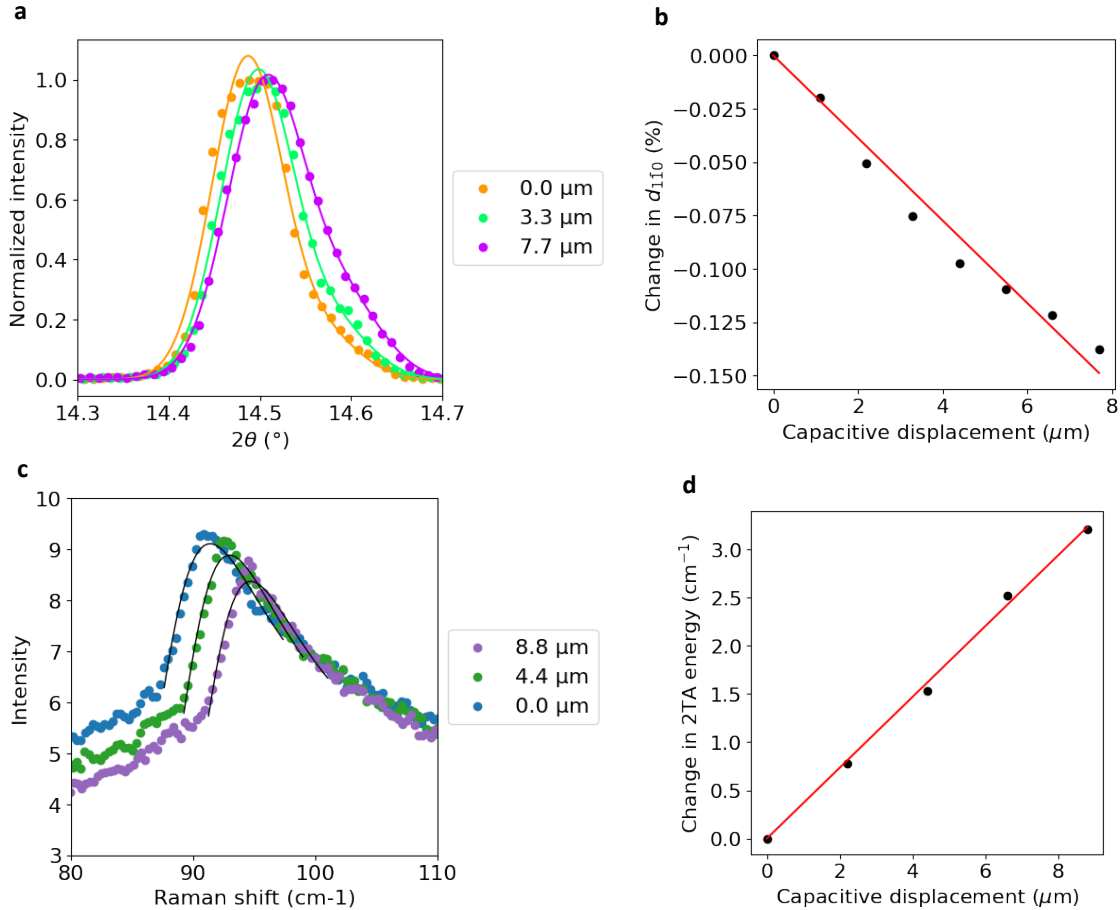

**Fig. S10**:(a) $\theta - 2\theta$ scans of the (1$\bar{1}$0) Bragg reflection measured at different capacitive displacement values. The solid lines are fits to a double-pseudovoigt function. (b) Percentage change in the distance between the (1$\bar{1}$0) lattice planes ($d_{1\bar{1}0}$) obtained from the shift of the (1$\bar{1}$0) Bragg reflection as a function of displacement. The solid line is the result of a linear fit to the data. (c) Raman spectra of the second-order TA phonon peak at ~90 cm$^{-1}$ measured at different displacement values. The solid lines are the results of fits to the data near the peak, using a Fano lineshape. (d) Energy of the second-order TA phonon at ~ 90 cm$^{-1}$ as a function of capacitive displacement. The solid line is the result of a linear fit to the data. All data were obtained at 30 K.

XRD measurements were performed using an in-house four-circle diffractometer equipped with a Xenocs Genix3D Mo K$\alpha$ (wavelengths 0.7093 and 0.7135 Å) x-ray source, with a beam spot diameter of 200 µm at the sample position. Figure S10(a) show $\theta - 2\theta$ scans across the [1$\bar{1}$0] Bragg peak at different capacitive displacements. Note that two peaks are visible due to the presence of two wavelengths in the incoming beam. The extracted spacing between the (1$\bar{1}$0) planes, $d_{1\bar{1}0}$, exhibits a linear relationship with the capacitive displacement, as shown in Fig. S9(b), with a slope of 0.019%/µm. As shown in Figs. S10(c,d), Raman scattering reveals a linear relationship between the capacitive displacement and the energy of the second-order TA peak at ~ 90 cm$^{-1}$, with a slope of 0.37 cm$^{-1}$/µm. By combining these results, the strain per unit shift of the second-order TA phonon mode is determined to be 0.019/0.37 = 0.051 %/cm$^{-1}$. For the conversion of the strain to stress values, we used an elastic modulus of 301 GPa, calculated for the [1$\bar{1}$0] axis from the elastic constants reported in [S5].

In all cases, the zero-strain reference value ("0 MPa") is determined from the unstrained portion of the sample that extends through the holes on either side of the sample carrier plates (see Fig. 1d in the main text).

## VII. Raman measurements for [001] stress on an OVD-STO sample deformed beyond the elastic regime

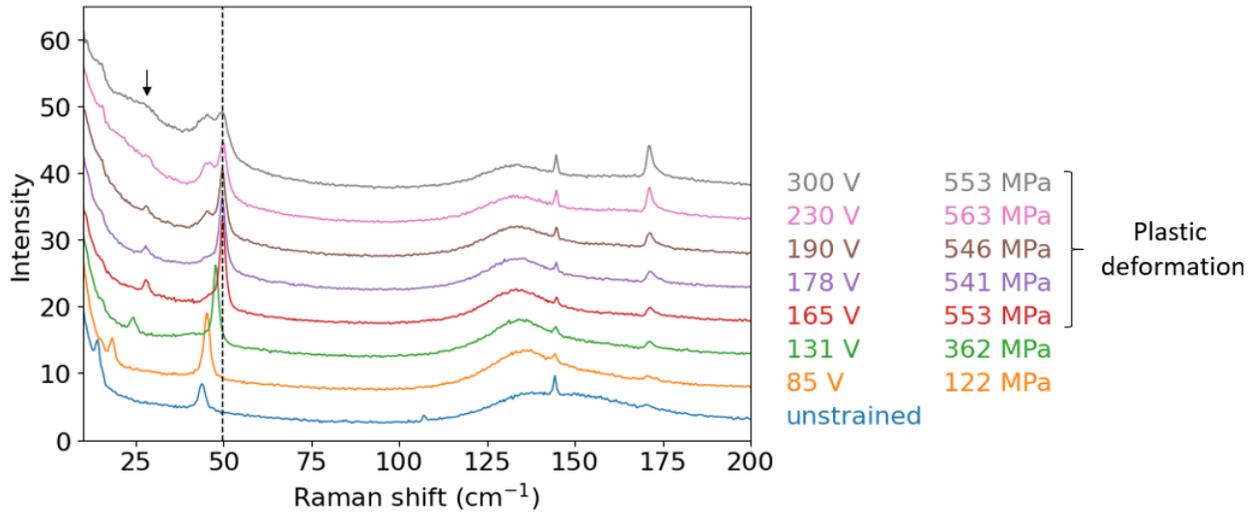

**Fig. S11**: Raman spectra of an OVD-STO sample strained along the [001] direction. The labels indicate the voltage applied to the piezoelectric stacks and the corresponding stress for each spectrum, extracted from the position of the $A_{1g}$ phonon near ~44 cm$^{-1}$. All spectra were obtained at 30 K.

Fig. S11 shows the Raman spectra of an OVD-STO sample strained along the [001] direction. Beyond an applied voltage of 165 V on the piezoelectric stacks, the $A_{1g}$ phonon mode—used to extract the stress—no longer shifts, indicating the onset of plastic deformation. This is further supported by the nearly constant stress values observed for data at and above 165 V. In this plastically deformed regime, an additional soft TO mode is observable as a distinct peak at ~27 cm$^{-1}$, in addition to the soft TO mode already visible in the elastic regime (see Fig. 3f of main text). The latter likely corresponds to the phonon mode measured with neutron scattering. Note that the feature at ~15 cm$^{-1}$ is an artefact from the optics in the Raman setup.

## VIII. Raman measurements for [001] stress on KTO

Fig. S12 shows the Raman spectra of a KTO sample strained along the [001] direction, measured at 30 K. The soft TO1 modes between 10 cm$^{-1}$ and 70 cm$^{-1}$ become Raman active between 652 MPa and 778 MPa (see Fig. S11(a,b)). Additionally, as shown in Fig. S12(c), the intensity of the TO2 mode (~197 cm$^{-1}$) also increases within this stress range. These results are in close agreement with previous work at 2 K [S5], where the TO modes become Raman active at ~700 MPa, indicating that the ferroelectric phase transition in KTO occurs at nearly the same stress at both 2 K and 30 K. This supports the validity of comparing neutron data obtained at 2 K with Raman data measured at 30 K.

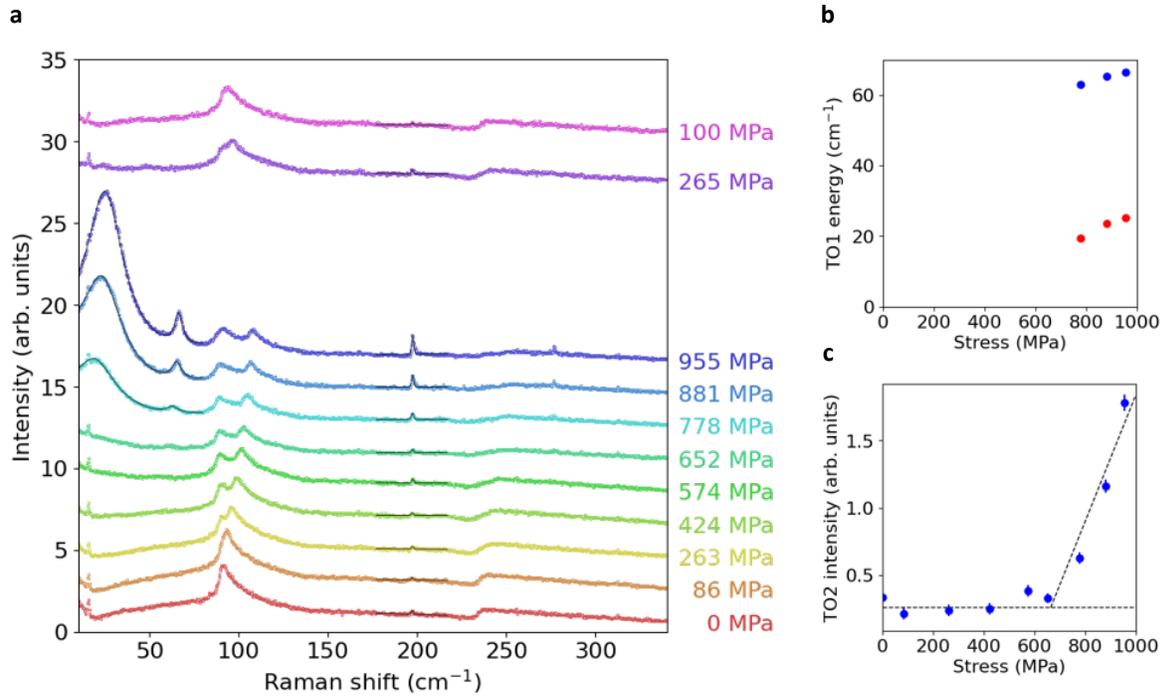

**Fig. S12**: (a) Raman spectra of a KTO sample strained along the [001] direction. The labels indicate the stress for each spectrum, extracted from the position of the second-order TA phonon at ~90 cm$^{-1}$ (see Fig. S9(b)). All spectra were obtained at 30 K. (b) Energy of the soft TO1 modes between 10 cm$^{-1}$ and 70 cm$^{-1}$ as a function of applied stress, extracted with a double-Lorentzian fit to the spectra (see solid lines in (a)). (c) Integrated intensity of the TO2 mode at ~197 cm$^{-1}$, extracted from resolution-limited Gaussian fits to the peak (see solid lines in (a)).

## IX. Additional density functional theory (DFT) results for KTO

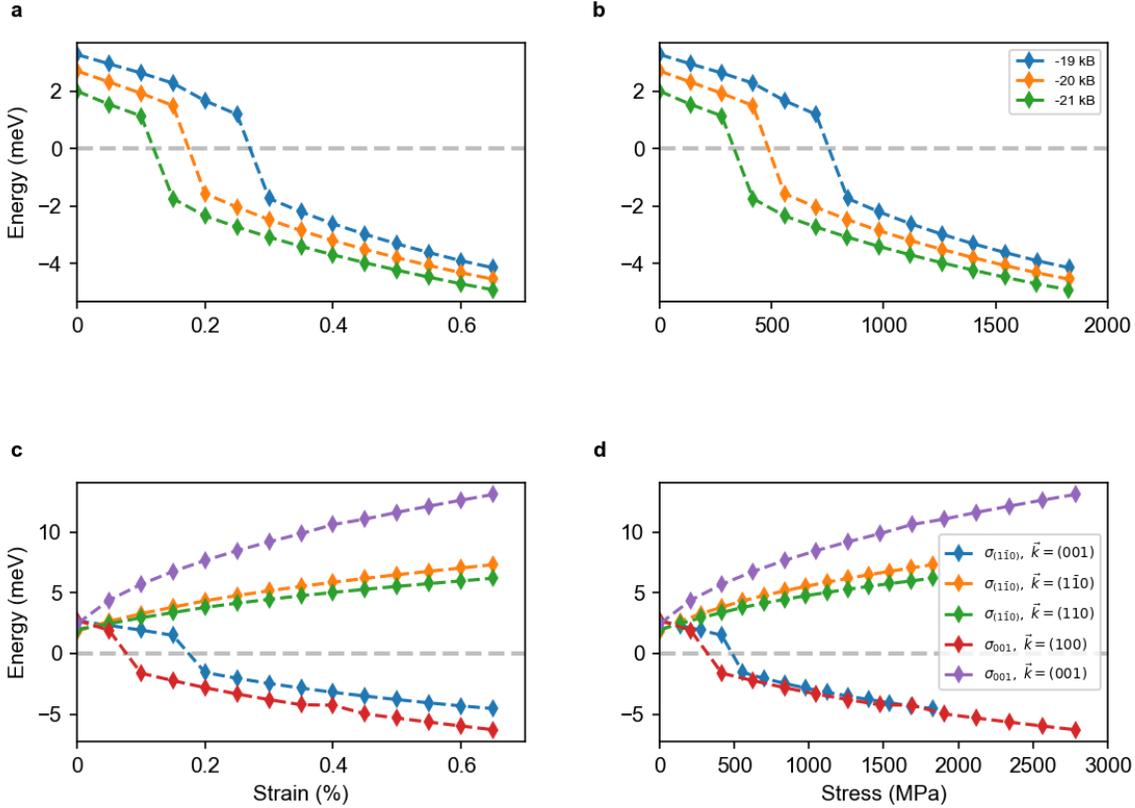

**Fig. S13:** Soft optical phonon frequencies for KTO strained along [001] and [1$\bar{1}$0] with spin-orbit coupling. (a, b) Phonon energies of the unstable mode for KTO strained along [1$\bar{1}$0] with hydrostatic pressures of -19, -20, and -21 kB as a function of strain and stress, respectively. After determining that a hydrostatic pressure of -20 kB causes the behavior of the polar phonon under strain to align closely with experiment, both the unstable and stable modes were calculated (c, d) Unstable and stable modes for KTO strained along [1$\bar{1}$0] and [001]. For [1$\bar{1}$0] strain, calculations find a single unstable phonon with displacement along (001). There are two singly degenerate stable modes; one that is along (110) and another along (1$\bar{1}$0).

**References**

[S1] Comes, R. & Shirane, G. Neutron scattering analysis of the linear-displacement correlations in $KTaO_3$. *Phys. Rev. B* **5**, 5 (1972).

[S2] Axe, J. D., Harada, J. & Shirane, G. Anomalous acoustic dispersion in centrosymmetric crystals with soft optic phonons. *Phys. Rev. B* **1**, 1227 (1970).

[S3] Uwe, H. & Sakudo, T. Stress-induced ferroelectricity and soft phonon modes in $SrTiO_3$. *Phys. Rev. B* **13**, 271 (1976).

[S4] Bouafia, H., Hiadsi, S., Abidri, B., Akriche, A., Ghalouci, L., Sahil, B. Structural, elastic, electronic and thermodynamic properties of $KTaO_3$ and $NaTaO_3$: *Ab initio* investigations. *Comput. Mat. Sci.* **75**, 1 (2013).

[S5] Uwe, H. & Sakudo, T. Raman-scattering study of stress-induced ferroelectricity in $KTaO_3$. *Phys. Rev. B* **15**, 1 (1977).